\documentclass[12pt,twoside]{article}
\usepackage{fleqn,espcrc1,amssymb}
\usepackage{graphicx}
\usepackage{epsfig}
\usepackage[figuresright]{rotating}

\newcommand{\AmS}{{\protect\the\textfont2
  A\kern-.1667em\lower.5ex\hbox{M}\kern-.125emS}}

\newcommand\numu{{\nu_\mu}}
\newcommand\anumu{\bar\nu_\mu}

\newcommand{\beq}{\begin{equation}}
\newcommand{\eeq}{\end{equation}}
\newcommand{\bea}{\begin{eqnarray}}
\newcommand{\eea}{\end{eqnarray}}
\def\dm2{\Delta m^2}
\def\sq2{sin^2(2\Theta)}
\def\nubar{\overline {\nu} }

\newcommand{\FLUKA}{{\sc FLUKA}}
\newcommand{\PEANUT}{{\sc PEANUT}}

\title{The \FLUKA{} atmospheric neutrino flux calculation}

\author{\footnotesize G.~Battistoni\address{INFN and Universit\`a di
Milano,  Dipartimento di Fisica, Milano, 20133, 
Italy}, A.~Ferrari
\address{CERN, Geneva 23, Switzerland,
on leave of absence from INFN, Milano.},
T.~Montaruli\address{Dipartimento di Fisica dell'Universit\`a e INFN, Bari},
P.R.~Sala
\address{ETH, Zurich, Switzerland,
on leave of absence from INFN, Milano.}
}
\begin{document}

\maketitle

\begin{abstract}
The 3--dimensional (3--D) calculation of the atmospheric neutrino flux by means of the \FLUKA{} Monte Carlo
model is here described in all details, starting from the latest data on primary cosmic ray spectra.
The importance of a 3--D calculation and of its consequences have been already debated
in a previous paper.
Here instead the focus is on the absolute flux. 
We stress the relevant aspects of the hadronic interaction model of \FLUKA{}
in the atmospheric neutrino flux calculation. 
%TM
This model is constructed
and maintained so to provide a high degree of 
accuracy in the description of %the properties of 
particle production.  The accuracy achieved in the comparison
with data from accelerators and cross checked with data
on particle production in atmosphere certifies
the reliability of shower calculation in atmosphere.

The results presented here can be already used for analysis by current
experiments on atmospheric neutrinos. 
However they represent an intermediate step towards
a final release, since this calculation does not yet include the
bending of charged particles in atmosphere. On the other hand this last
aspect, while requiring a considerable effort in a fully 3--D description
of the Earth, if a high level of accuracy has to be 
maintained, does not affect in a significant way  
the analysis of atmospheric neutrino events.
\end{abstract}

\section{Introduction}
\label{sec:intro}
Reliable calculations of flux of secondary particles in atmosphere, produced
by the interactions of primary cosmic rays, are essential
for the correct interpretation of the large amount of
experimental data produced by experiments in the field of
astroparticle physics. The increasing accuracy of modern
experiments demands also an improved quality of the calculation tools.
The most important example in this field is the analysis of 
the experimental results on atmospheric neutrinos from 
Super--Kamiokande\cite{superk}, MACRO\cite{macro} and
Soudan2\cite{soudan2}, which gave the first robust evidence
in favor of neutrino oscillations. The interpretation in
terms of the mixing parameters is affected by different sources of
systematic errors, and the theoretical uncertainties on neutrino
fluxes and cross sections constitute a significant fraction of them.
This has stimulated different efforts to improve the existing 
flux calculations. Among them, one is the calculation based on the \FLUKA{}
Monte Carlo code\cite{fluka}. Such a work was indeed started
before\cite{taup97} 
the Super--Kamiokande results, in the framework of design work for the ICARUS
experiment\cite{icarus} and of the analysis of MACRO experiment at Gran
Sasso. 
The main motivation to propose the 
%TM
\FLUKA{} based calculation 
was the idea that, in order to accomplish the goals 
summarized before, 
the highest degree of detail and accuracy should be accomplished.

In particular the
\FLUKA{} code is known for the accuracy of its particle production
model in hadronic interactions, which 
is extensively benchmarked against accelerator data.
The first important achievement obtained using
\FLUKA{} concerned the relevance of 3--D geometry
in the flux calculations and
was presented in ref.\cite{flukanu}.
However, at the time of that work it was not yet possible to
make statements on the absolute values of neutrino fluxes. 
This is instead the purpose of the
present paper, which is intended to provide a complete reference for
the \FLUKA{} neutrino flux.

In the following, we shall first summarize the main ingredients of
the simulation, discussing with some detail the 
primary cosmic ray spectrum and the geometrical set--up. 
Then, in a next section, we shall discuss
the physics models of \FLUKA{} presenting a set of comparisons between
data and predictions to demonstrate the validity of the models themselves.
The final results on 
the flux are presented and discussed,
also by means of a comparison to other calculation results.
Neutrino fluxes are calculated for 3 
relevant experimental sites (Kamiokande, Gran Sasso and Soudan) and 
representative tables are given here for the Kamiokande site.
More detailed tables for all the 3 sites are
available on the web\cite{flukatab}.
The energy region considered in this work (0.1$\div$200 GeV) 
covers in practice the production of the event topologies of 
 of sub-GeV and multi-GeV events as detected in Super--Kamiokande.

In the conclusions we shall debate the level of systematic
error in  the predictions and mention the items that will require
further improvements.

\section{The Simulation Set--up}
\label{sec:setup}
The flux calculation has been completely carried out in the framework
of the \FLUKA{} Monte Carlo code\cite{fluka}.
This is an interaction and transport 
MonteCarlo package able to treat with a high degree of detail 
the following problems:

\begin{itemize}
\item Hadron-hadron and hadron-nucleus interactions (0-100~TeV);
\item Electromagnetic and $\mu$ interactions (1~keV-100~TeV);
\item Charged particle transport - ionization energy loss;
\item Low energy neutron multigroup transport and interactions (0-20~MeV)
\end{itemize}

In the following we present some detail about the essential elements
of the calculation, {\it i.e.} the geometrical description of the problem , the
primary spectrum, the geomagnetic model and the treatment of solar modulation.

\subsection{Simulation geometry, description of Earth and of its atmosphere}
\begin{figure}[htbp]
\begin{center}
\makebox[150mm]{\epsfig{file=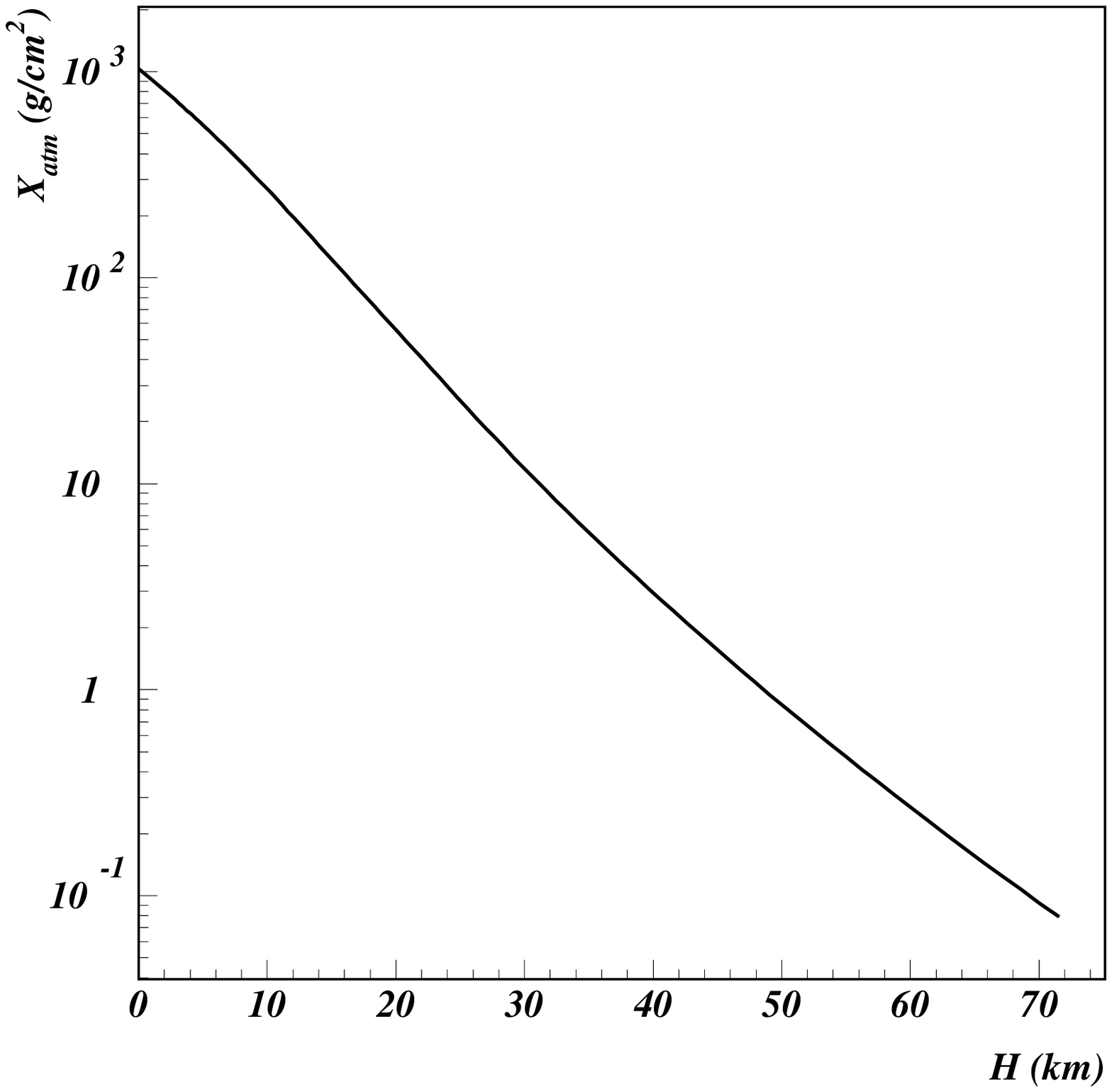,width=150mm}}
\caption{Vertical atmospheric depth as a function
of height above sea level as used in the FLUKA simulation.
\label{fig:atmo}
}
\end{center}
\end{figure}

In view of applications for cosmic ray physics, we have implemented a
3--D 
spherical representation of the whole 
%TM
Earth 
and of the surrounding atmosphere.
The radius of the Earth sphere is assumed to be $R~=~6378.14$ km.
The atmosphere is described by a medium composed
by a proper mixture of N, O and Ar, arranged in 100 concentric
spherical shells, each one having a density scaling according to the
known profile of ``standard atmosphere" \cite{atmo} (see
Fig.~\ref{fig:atmo}).
 This can be considered an improvement with respect to the work of
ref.\cite{flukanu}, 
where only 51 shells where considered. However, we
could not find any significant changes in all test runs that we performed
comparing results for the two different realizations of the atmosphere.

%TM More important is the fact that, (toglierei importanza alla cosa) 
It should be noticed that, as in the other standard neutrino
calculations, we are using the same description of the atmosphere profile
over the whole Earth, while it is known that there exist differences as a
function of latitude\cite{cospar} and as a function of time during the
year. This in principle is a source of systematic error, that will be
discussed in section \ref{sec:syserr}.

\subsection{The Primary C.R. spectrum}
\label{sec:prim}
In the first reference of FLUKA neutrino flux, the original 
%TM
Bartol spectrum of all nucleon flux was used\cite{bartol}.
The calculation presented here makes use of the fit of the recent cosmic
ray measurements presented in
\cite{bartol_new} 
to recent cosmic ray flux measurements.
In addition, the authors of \cite{bartol_new} take into account the
statistical errors of measurements and the systematic differences
between different experiments in order to establish a band of uncertainty
around the average result of this new fit. 

The functional form adopted by Bartol to fit the data point as a function of 
the kinetic energy per nucleon ($E_k$) is:
\begin{equation}
\phi (E_k) = K \times \left( E_k + b\  exp\left[-c\sqrt E_k \right] \right)^{-\gamma}
\end{equation}
In Table \ref{tabnum} we report the values of the four fit parameters ($K$,
$b$, $c$ and $\gamma$) for the different mass groups considered by Bartol.
The higher and lower extremes for the proton components are obtained
by considering only the error on $K$. The central value for the nuclear
components is obtained by the average of the ``High'' and ``Low''
parameter values.
 
\begin{table}[thb]
 \begin{center}
  \caption {Fit parameters 
%TM
from Ref.~\protect\cite{bartol_new} for 
the main 
%TM all 
five components of the primary cosmic rays \label{tabnum}}
 \vspace*{2truemm}
  \begin{tabular}{|l|l| l l l l|} \hline
  component/fit & $<A>$ & $\gamma$ &  K
& b & c  \\ 
    &  & & \it{cm$^{-2}$ s$^{-1}$ sr$^1$
 (GeV/A)$^{-1}$}
&\it{GeV/A} & \it{(GeV/A)$^{-0.5}$} \\ 
\hline
 protons  & 1 & 2.74$\pm$0.01 & 14900$\pm$600 & 2.15 & 0.21 \\
 He (high) & 4 & 2.64$\pm$0.01 & 600$\pm$30  & 1.25 & 0.14 \\
 He (low)  & 4 & 2.74$\pm$0.03 & 750$\pm$100 & 1.50 & 0.30 \\
 CNO (high) & 14 & 2.70      & 68          & 1.78 & 0.02 \\
 CNO (low)  & 14 & " "       & 55          & " " & " " \\
 MgSi (high) & 26 & " "      & 22          & " " & " " \\
 MgSi (low)  & 26 & " "      & 18          & " " & " " \\
 $z>$17(high)& 50 & " "      & 5.50        & " " & " " \\
 $z>$17(low))& 50 & " "      & 5.20        & " " & " " \\ \hline
  \end{tabular}
 \end{center}
\end{table}

Primary particles, sampled
from such continuous spectra, are injected
at the top of the atmosphere, 
at about 100 km of altitude. The primary flux is assumed to be 
uniform and isotropic. In case of primary nuclei, 
individual nucleons
are unpacked before interaction, according to the 
the superposition model, since the present version of \FLUKA{} code
does not yet handle nuclear projectiles (see section \ref{sec:hadro}).

The flux of all possible secondary products are
scored at different heights in the atmosphere 
and at the earth boundary. 
Therefore,
we are able to get, besides neutrino fluxes,
the flux of muons and hadrons to be used for the benchmarking against
existing experimental data.

\subsection{The Geomagnetic Field}
\label{sec:geomag}

For our calculation we have adopted the IGRF model for geomagnetic field,
which is expressed as an expansion of spherical harmonics\cite{igrf}.
A complete calculation within the 3--D Earth geometry
should include geomagnetic effects both to introduce primary cutoff
and to include charged particle bending during shower
development\cite{lipgeo}. 
The cutoff must be explicitly
computed for each cosmic ray taken into consideration. In principle, one
should consider primary cosmic rays from all directions over the whole
Earth surface, and then, after applying the cutoff, count the neutrino
reaching an area around the location of interest sufficiently small
to avoid any significant change in the geomagnetic cutoff ({\it i.e.} few
square degrees or less). 
This procedure does not allow to exploit
the spherical symmetry to minimize computer power, since a realistic
description of the magnetic field makes each point different from the
other. However, the solutions so far adopted in ref. \cite{lipgeo} and
\cite{hondanew} are still unsatisfactory.
In ref.\cite{hondanew} the authors make use of a
dipole field both for geomagnetic cutoff test and
particle simulation in atmosphere.
While we agree on the use of a dipole field for
shower development, we consider this as a too rough approximation
for cutoff evaluation purposes.
In ref.\cite{lipgeo}, which was meant to be mostly a demonstrative work and
not a flux reference, a realistic description of geomagnetic field
is chosen, but in order to make the calculus viable, ``detector regions''
were defined as large portions of the Earth's surface. 

We could not yet find a satisfactory solution, in terms of
calculation power and precision at the same time, to include
bending of charged particles during shower development. 
Therefore we have generated particles neglecting this aspect,
while the field is accounted to determine the cutoff conditions 
for primary particles as seen by a specific detector location.
This is performed according to the following procedure,
which maximizes the use of the available information:

\begin{itemize}
\item neutrinos reaching the earth surface have been recorded with their
impact position, energy, direction and  
%TM 
with their primary cosmic ray
position, energy, direction,
charge and mass at the point of the first
interaction in the atmosphere. No geomagnetic effect has been applied at
this stage.
\item Offline, all recorded neutrinos and their parents have been rotated
in such a way to make their impact position correspond to the selected
detector site. At the same time  the corresponding upward going neutrino
(and the position and direction of its parent primary cosmic ray) has 
been created out of
simple geometrical considerations.  Note please that at this stage 
the problem is still fully symmetric and each downward going neutrino has a
corresponding upward going one 
%TM in
at the same location.
\item The geomagnetic cutoff has been applied using standard techniques, 
making use of the information about the relative position and direction of
the parent primary cosmic ray.

\end{itemize}

In this way we have been able to exploit all available statistics for
each relevant detector site. On the other hand we are aware that 
neglecting bending of charged particles in atmosphere introduces
some systematic errors. According to ref.\cite{lipgeo} these should
manifest themselves in particular observables, such as, for example,
the East--West asymmetry. However it can be shown that we are talking
of effects at the level of percent, which hardly affect the analysis of present neutrino
events in terms of new physics. Therefore we prefer, for the moment,
to account for this approximation in our systematic error budget 
(see section \ref{sec:syserr}
%TM
).

\subsection{Solar Modulation}
The cosmic ray spectrum depends on the phase of solar
cycle. In order to modulate our primary spectrum, we adopt an
algorithm proposed in ref.\cite{modulation}. There the effect
of solar modulation is treated in the framework of the
``force field approximation''\cite{gleason} which eventually
express the modification of the spectrum as a function of energy
as if due to an effective potential. 
For a given distance from the sun, this potential is a function
only of time, and this has been parametrized as a function
of the counting rate of the Climax neutron monitor\cite{climax},
which has worked until year 2000. In this way we have evaluated
the two extreme cases corresponding to average ``minimum'' and
``maximum'' solar activity.
The algorithm adopted here is independent from those adopted in
other authors' works like those of ref. \cite{hkkm} and \cite{bartol}.

\section{The Description of Hadronic Interactions}
\label{sec:hadro}
The hadronic packages of the \FLUKA{} Monte Code are being developed
according to a theoretically inspired approach, namely that interactions 
should be described in terms of the properties of elementary constituents. 
In principle one would like to derive all features of ``soft" 
interactions (low-$p_T$ interactions) from the QCD Lagrangian, as it is 
done for hard processes. Unfortunately the large value taken by the 
running coupling constant prevents the use of the perturbative calculus. 
Indeed, in QCD, the color field acting among quarks is carried by the
vector bosons of the strong interaction, the gluons, which are
``colored" themselves. Therefore the characteristic feature of gluons
(and QCD) is their strong self-interaction. If
we imagine that quarks are held together by color lines of force, the
gluon-gluon interaction will pull them together into the form of a tube
or a string. Since quarks are confined, the energy required to
``stretch" such a string is increasingly large until it suffices to
materialize a quark-antiquark couple from the vacuum and the string
breaks into two shorter ones, with still quarks at both ends.
Therefore it is not unnatural that, because of quark confinement, theories
based on interacting strings emerged as a powerful tool in understanding
QCD at the soft hadronic scale (the non-perturbative regime).
Different implementations of this idea exist, having obtained
remarkable success in describing the features of hadronic interactions.
These concepts are embedded in \FLUKA{} in order to give sound
physical bases to each step.

It must be remarked that it is not yet possible to cover all relevant 
energy ranges by means of a unique numerical model. Different 
approaches are anyway required, and this imposes particular care so to
have the proper matching and continuity at transition energies.
Of course this point has been analysed in detail by \FLUKA{} authors.
In any case, a part the attention to the phenomenological aspects, 
the performances of the code are then
optimized comparing with particle production data at
single interaction level. 
The final predictions are obtained with a minimal set of free parameters,
fixed for all energies and target/projectile combinations.
Results in complex cases as well as scaling laws and properties come out
naturally from the underlying physical models. The basic conservation
laws are fulfilled ``a priori''.
This kind of approach has guaranteed a very high level of
detail in \FLUKA{} at least in principle. This is the reason why we
propose its use when
%TM when
aiming at precision calculations
with the idea of reaching a deeper understanding and reliability of
predictions.

\FLUKA{} contains two models to describe nonelastic hadronic interactions.
\begin{itemize}
\item The ``low-intermediate'' energy one, called \PEANUT{}, which covers
the energy range up to 5~GeV.
\item The high energy one which can be used up to several tens of TeV,
based on the color strings concepts sketched above.
\end{itemize}

The nuclear physics embedded in the two models is very much the same.
The main differences are a coarser nuclear description (and no preequilibrium
stage) and the Gribov-Glauber cascade for the
high energy one.

\subsection{The \PEANUT{} Model}
\label{sec:peanut}
Hadron-nucleus non-elastic interactions are often described
in the framework of the IntraNuclear Cascade (INC) models.
%TM
These kind of models was 
developed at the very beginning of the history of energetic nuclear 
interaction modelling, but it is still valid and in some energy range 
it is the only available choice.
Classical INC codes were based on a more or less accurate treatment of 
hadron multiple collision processes in nuclei, the target being assumed 
to be a cold Fermi gas of nucleons in their potential well.
The hadron-nucleon cross sections used in the calculations are 
free hadron--nucleon cross 
sections. Usually, the only quantum mechanical concept incorporated was 
the Pauli principle. Possible hadrons were often limited to pions and 
nucleons, pions being also 
%TM
produced 
or absorbed via isobar (mainly 
$\Delta_{33}$) formation, decay, and capture. 
Most of the historical weaknesses of INC codes have been mitigated or 
even completely solved in some of the most recent 
developments~\cite{FLUKA1,MASHNIK}, thanks to the inclusion of a 
so called ``preequilibrium'' stage, and to further quantistic effects including
coherence and multibody effects. 

All these improvements are considered in the \PEANUT{} 
(PreEquilibrium Approach to NUclear Thermalization) model of FLUKA.
Here the reaction mechanism is modelled in by explicit
intranuclear cascade smoothly joined to statistical (exciton)
preequilibrium emission~\cite{Gad92} and followed by evaporation 
(or fission or Fermi break-up) and gamma deexcitation.
In both stages, INC and exciton, the nucleus is modelled as a sphere
with density given by a symmetrized Woods-Saxon~\cite{Gry91}
shape with parameters according to the droplet model~\cite{Myers} for A$>$16, 
and by a harmonic oscillator shell model for light
isotopes (see~\cite{Elton}). 
The effects of the nuclear and Coulomb potentials outside the nuclear 
boundary are included. Proton and neutron densities are generally different.
%TM
Binding energies are obtained from mass tables.
Relativistic kinematics is applied at all stages, with
accurate conservation of energy and momentum including those of the residual
nucleus. Further details and validations can be found
in~\cite{FLUKA1}.

For energies in excess of few hundreds MeV the inelastic channels
(pion production channels) start to play a major role.
The isobar model easily accommodates multiple pion production, for 
example allowing the presence of more than one resonance in the 
intermediate state (double pion production opens already at 600~MeV 
in nucleon-nucleon reactions, and at about 350~MeV in pion-nucleon ones).
Resonances which appear in the intermediate states can be treated as real 
particles, that is, they can be transported and then 
transformed into secondaries according to their lifetimes and decay 
branching ratios. 

\subsection{The Dual Parton Model for high energy}
\label{sec:dpm}
A theory of interacting strings can be managed by means of 
the Reggeon-Pomeron calculus in the 
framework of perturbative Reggeon Field Theory\cite{Collins},
an expansion already developed before the 
establishment of QCD.
Regge theory makes use explicitly of the constraints of analyticity and 
duality. 
On the basis of these concepts, calculable models can be constructed
and one of the most successful attempts in this field is the so called 
``Dual Parton Model'' (DPM), originally developed in Orsay in
1979~\cite{DPMORI}.  
It provides the theoretical framework 
to describe hadron-nucleon interaction from several
GeV onwards. 
DPM exhibited remarkable successes in 
predicting experimental observables. The quoted references include a 
vast amount of material showing the capabilities of the model when 
compared with experimental data. However,
it must be stressed that other models are available, but most of them share an 
approach based on string formation and decay. For example, the {\it 
Quark Gluon String Model}~\cite{QGSM} has been developed more or less in
parallel with DPM, sharing most of the basic features of it.

In DPM a hadron is a low-lying excitation of an open
string with quarks, antiquarks or diquarks sitting at its ends. In
particular mesons  are
described as strings with their valence quark and antiquark at the
ends. (Anti)baryons are treated like open strings with a (anti)quark and a
(anti)diquark at the ends, made up with their valence quarks.

At sufficiently high energies,
the leading term in high energy scattering 
corresponds to a ``Pomeron''  exchange (a closed string exchange
with the quantum numbers of vacuum), 
which has a cylinder topology. By means of the optical theorem, connecting
the forward elastic scattering amplitude to the total inelastic cross
section, it can be shown that 
from the Pomeron topology it follows that two hadronic 
chains are left as the sources of particle production
(unitarity cut of the Pomeron). 
While the partons 
(quarks or diquarks) out of which chains are stretched carry a net 
color, the chains themselves are built in such a way to carry no net 
color, or to be more exact to constitute color singlets like all 
naturally occuring hadrons. In practice, as a consequence of color
exchange in the interaction, each colliding hadron splits into two
colored system, one carrying color charge $c$ and the other $\bar c$.
These two systems carry together the whole momentum of the hadron. The
system with color charge $c$ ($\bar c$) of one hadron combines with the
system of complementary color of the other hadron, in such a way to
form two color neutral chains. These chains appear as two back-to-back
jets in their own centre-of-mass systems.
The exact way of building up these chains depends on the nature of the
projectile-target combination (baryon-baryon, meson-baryon,
antibaryon-baryon, meson-meson).

The single Pomeron exchange diagram is the dominant contribution,
however higher order contributions with multi-Pomeron exchanges become
important at energies in excess of 1~TeV in the laboratory. They
correspond to more complicated topologies, and DPM provides a way for
evaluating the weight of each, keeping into account the unitarity
constraint. Every extra Pomeron exchanged gives rise to two
extra chains which are built using two $q\bar q$ couples excited from
the projectile and target hadron sea respectively. The inclusion of 
these higher order diagrams is usually referred  to as {\it multiple 
soft collisions}.

Two more ingredients are required to completely settle the problem. The 
former is the momentum distribution for the $x$ variables of valence 
and sea quarks. Although the exact form of the momentum distribution
function, $P(x_1,..,x_n)$, is not known, general considerations based on
Regge arguments allow one to predict the asymptotic behavior of this 
distribution whenever each of its arguments goes to zero. The behavior 
turns out to be singular in all cases, but for the diquarks. A 
reasonable assumption, always made in practice, is therefore to 
approximate the true unknown distribution function with the product of 
all these asymptotic behaviors, treating all the rest as a 
normalization constant.

The latter ingredient is a hadronization model, which must 
take care of transforming each chain into a sequence of physical 
hadrons, stable ones or resonances. The basic assumption is that of
{\it chain universality}, which assumes that once the chain ends and the 
invariant mass of the chain are given, the hadronization properties are 
the same regardless of the physical process which originated the chain.
Therefore the knowledge coming from hard processes and $e^+e^-$ 
collisions about hadronization can be used to fulfill this task. There 
are many more or less phenomenological models which have been developed 
to describe hadronization (examples can be found 
in~\cite{JETSET,BAMJET}). In principle hadronization properties too can 
be derived from Regge formalism~\cite{Kaifrag1}.

It is possible to extend DPM to hadron-nucleus collisions 
too~\cite{DPMORI}, making use of the so called Glauber-Gribov approach. 
Furthermore DPM provides a theoretical framework for
describing hadron diffractive scattering both in hadron-hadron and 
hadron-nucleus collisions. General information on diffraction in DPM
can be found in~\cite{GOULIA} and details as well as practical 
implementations in the DPM framework in~\cite{HANDIF1}.

At very high energies, those of interest for high energy cosmic ray studies 
(10--10$^5$~TeV in the lab), hard processes can no longer be
ignored. They are calculable by means of perturbative QCD and
can be included in DPM through proper unitarization 
schemes which consistently treat soft and hard processes together.

\subsection{Benchmarking of \FLUKA{} hadronic models with experimental data}
\label{sec:bench}

The predictions of \FLUKA{} have been checked with a large set of
experimental data collected in accelerator experiments. 
Reviews of these comparison with data can be found in \cite{cita}.
Here we shall limit ourselves to show only a few
examples, among the most important in view of the application of
the code to cosmic ray applications.

\begin{figure}[htbp]
\begin{center}
\makebox[150mm]{\epsfig{file=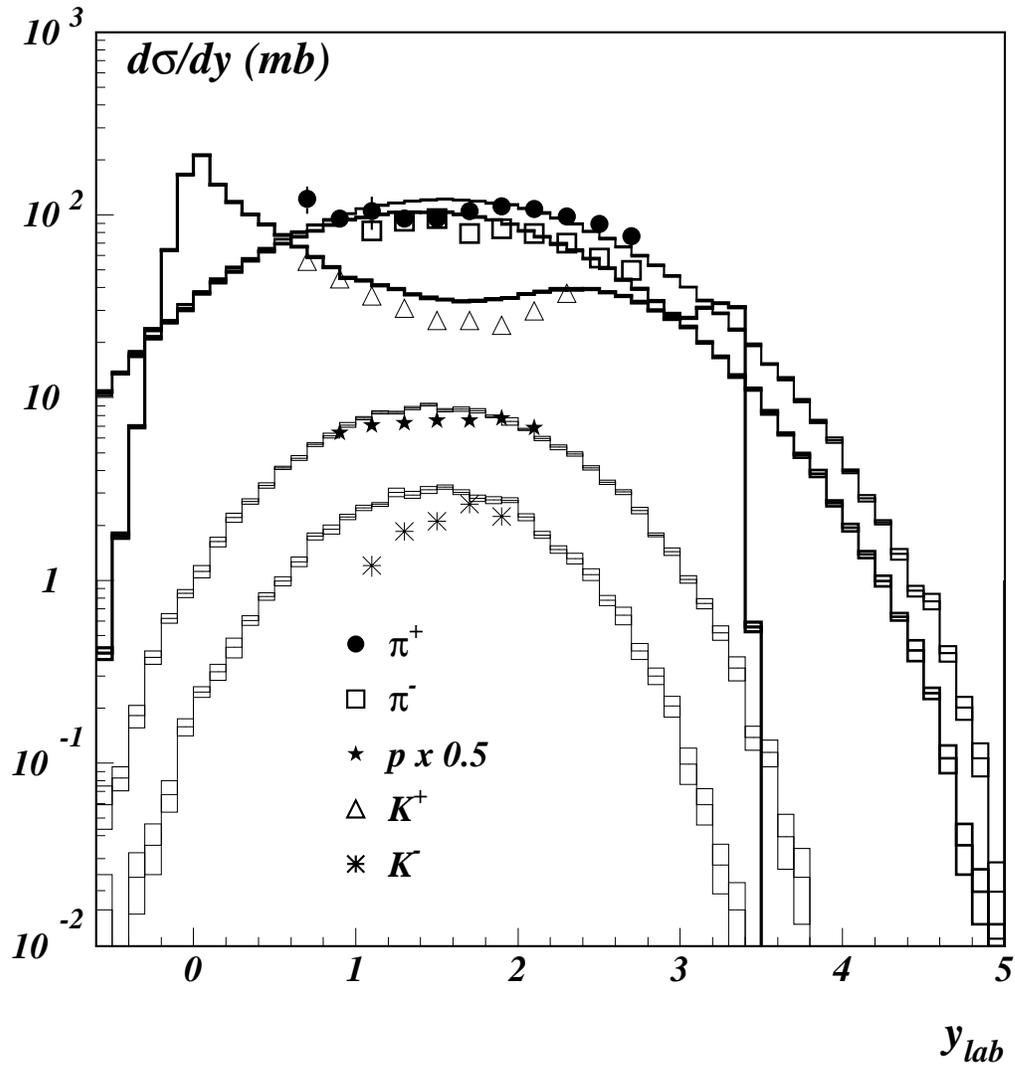,width=150mm%
,bbllx=60pt,bblly=100pt,bburx=590pt,bbury=740pt}} \\
\caption{
%TM
Rapidity distribution of $\pi^{+/-}$, K$^{+/-}$ and protons
for 14.6~GeV/c protons on Be (data from ref.~\protect\cite{abbott}).
Histograms are simulated data.
\label{fig:pbey}
}
\end{center}
\end{figure}

\begin{figure}[htbp]
\begin{center}
\begin{tabular}{cc}
\makebox[75mm]{\epsfig{file=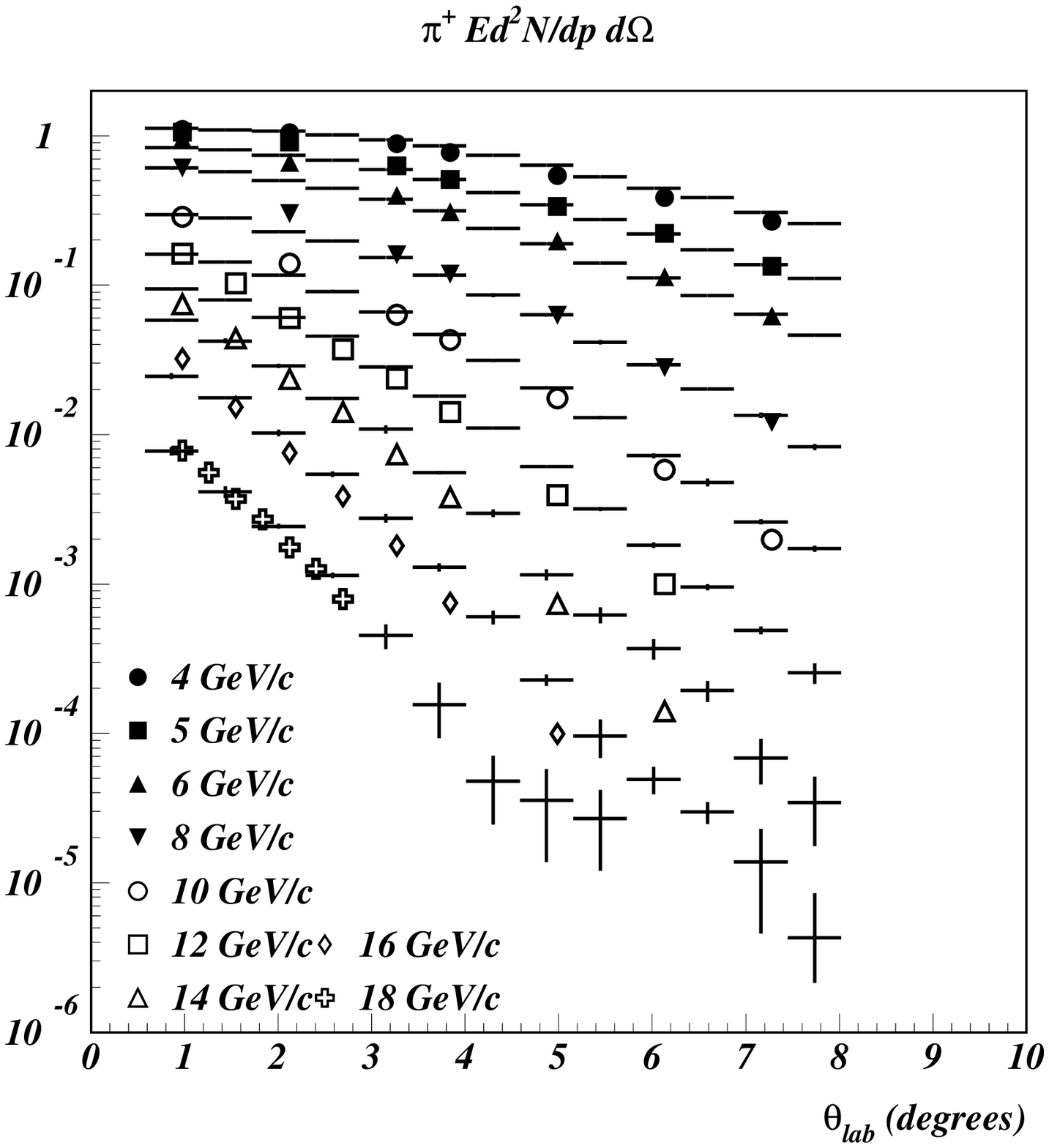,width=75mm}}&
\makebox[75mm]{\epsfig{file=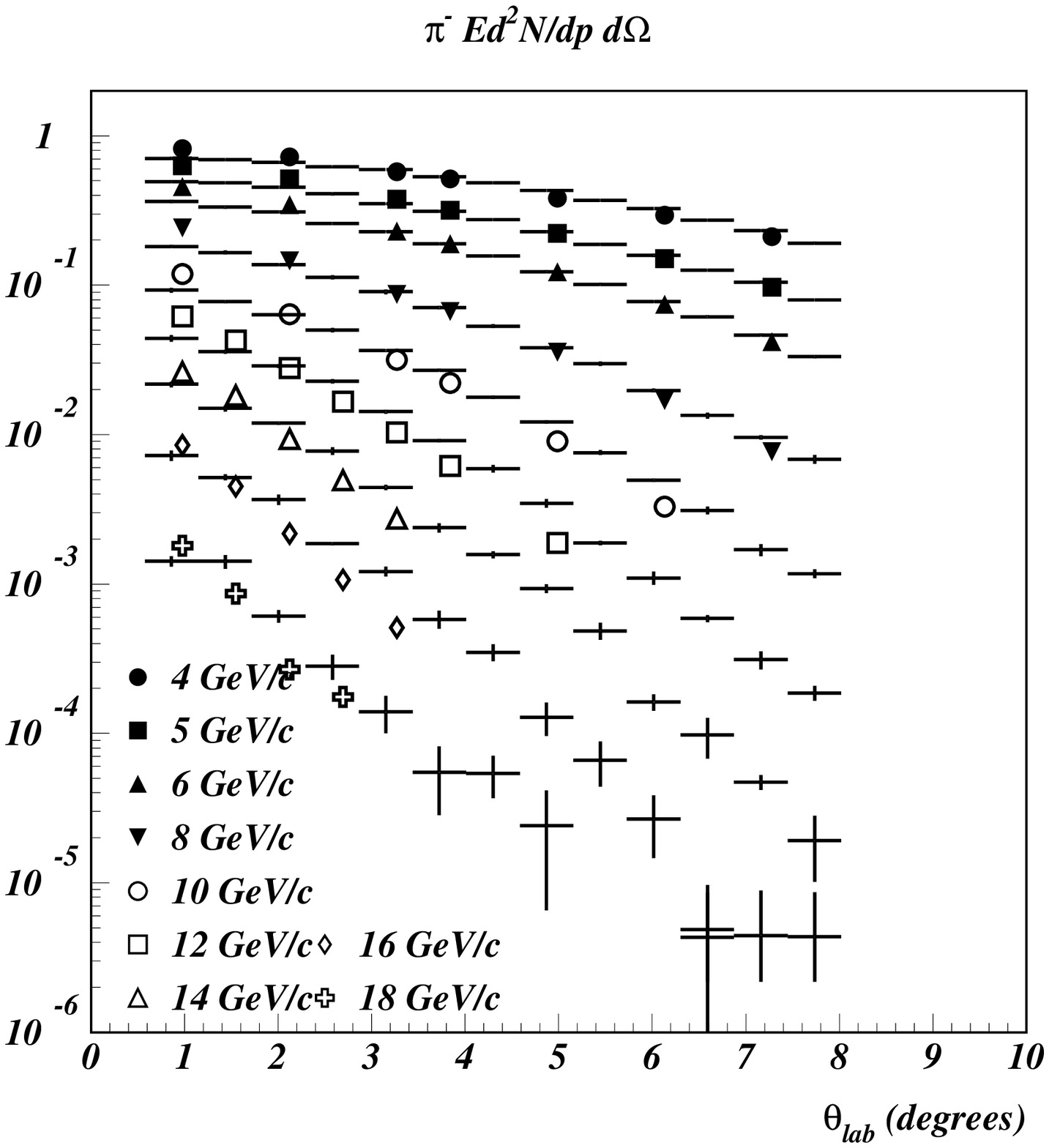,width=75mm}} \\
\end{tabular}
\caption{
Double differential particle yield
for $\pi^{+}$ (left) and $\pi^{-}$ (right) for 24~GeV/c 
protons on Be (symbols extrapolated from the double differential
cross section reported in ref.~\protect\cite{eichten}).
Histograms are simulation results.
Data are given as a function
of $\theta_{lab}$ for different momentum bins.
\label{fig:eich}
}
% }
\end{center}
\end{figure}

\begin{figure}[htbp]
\begin{center}
\begin{tabular}{cc}
\makebox[75mm]{\epsfig{file=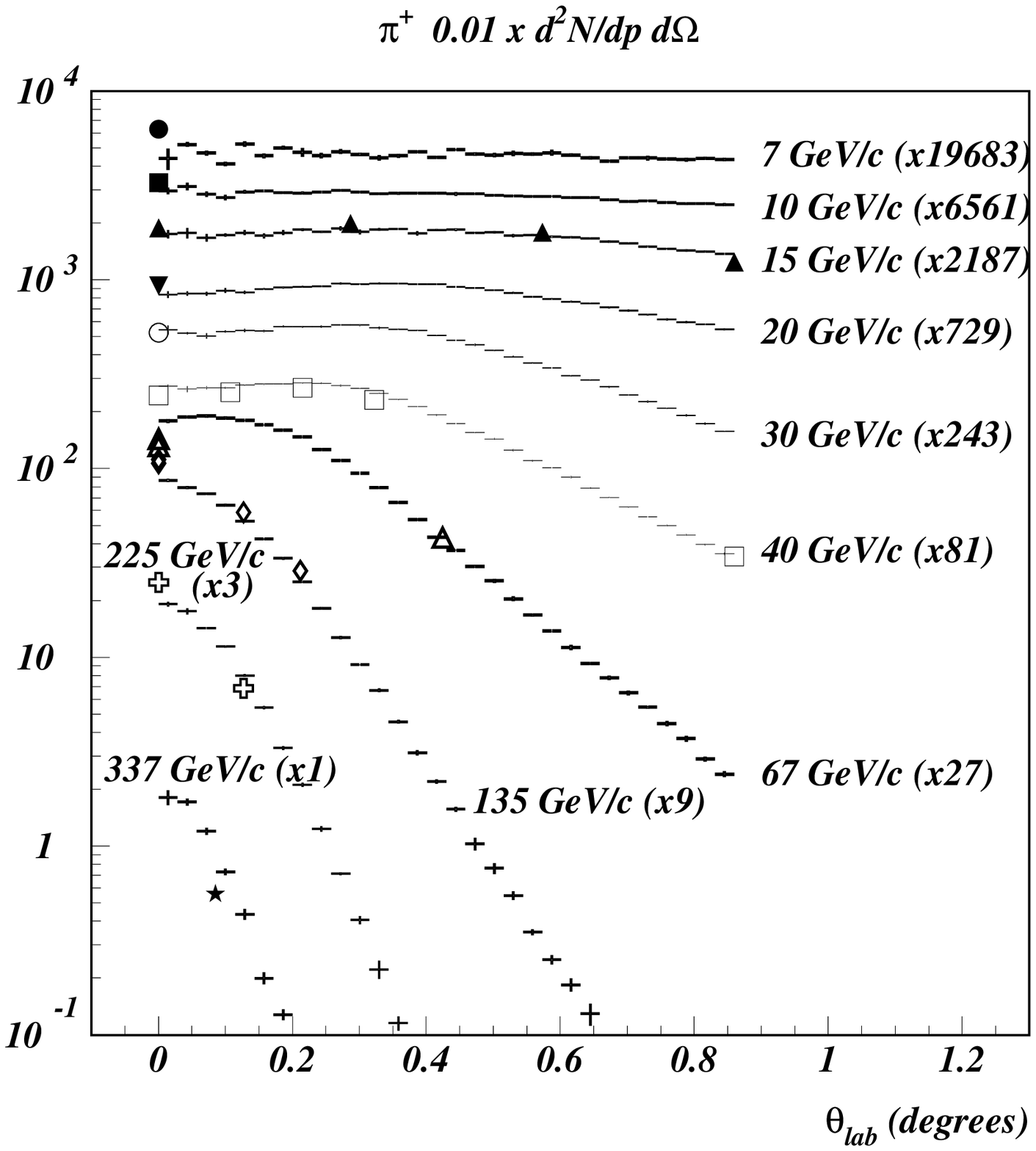,width=75mm}} &
\makebox[75mm]{\epsfig{file=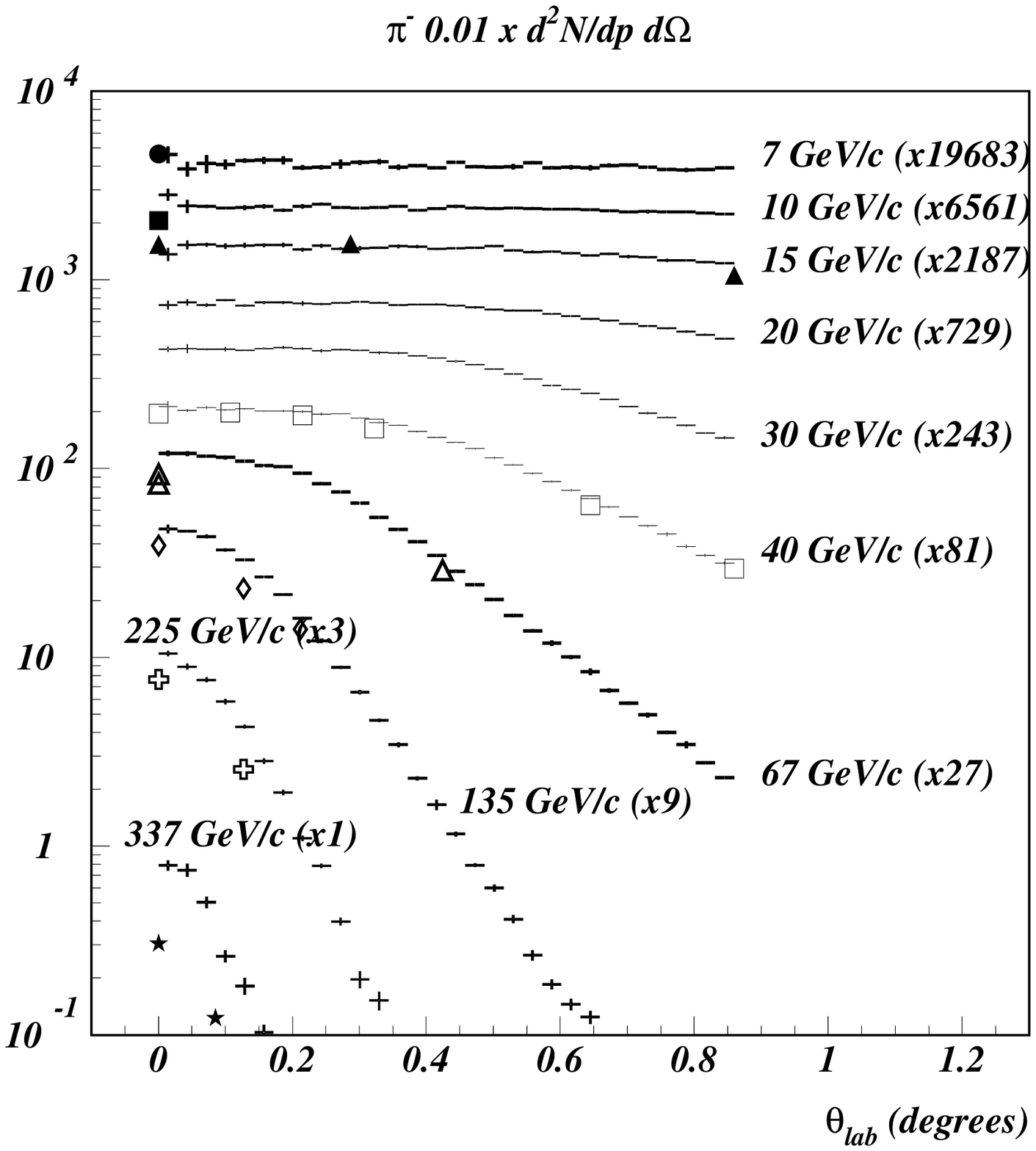,width=75mm}} \\
\end{tabular}
\caption{
Double differential cross section for 
$\pi^+$ (left) and $\pi^-$ (right)
production for 450~GeV/c protons on a 10 cm thick Be target 
(data from ref.~\protect\cite{atherton} and \protect\cite{spy}).
Histograms are simulation results. Data are given as a function
of $\theta_{lab}$ for different momentum bins. The
distributions are scaled by increasing factors (reported in the
figure) to allow a clearer separation.
\label{fig:spypi}}
\end{center}
\end{figure}

\begin{figure}[htbp]
\begin{center}
\begin{tabular}{cc}
\makebox[75mm]{\epsfig{file=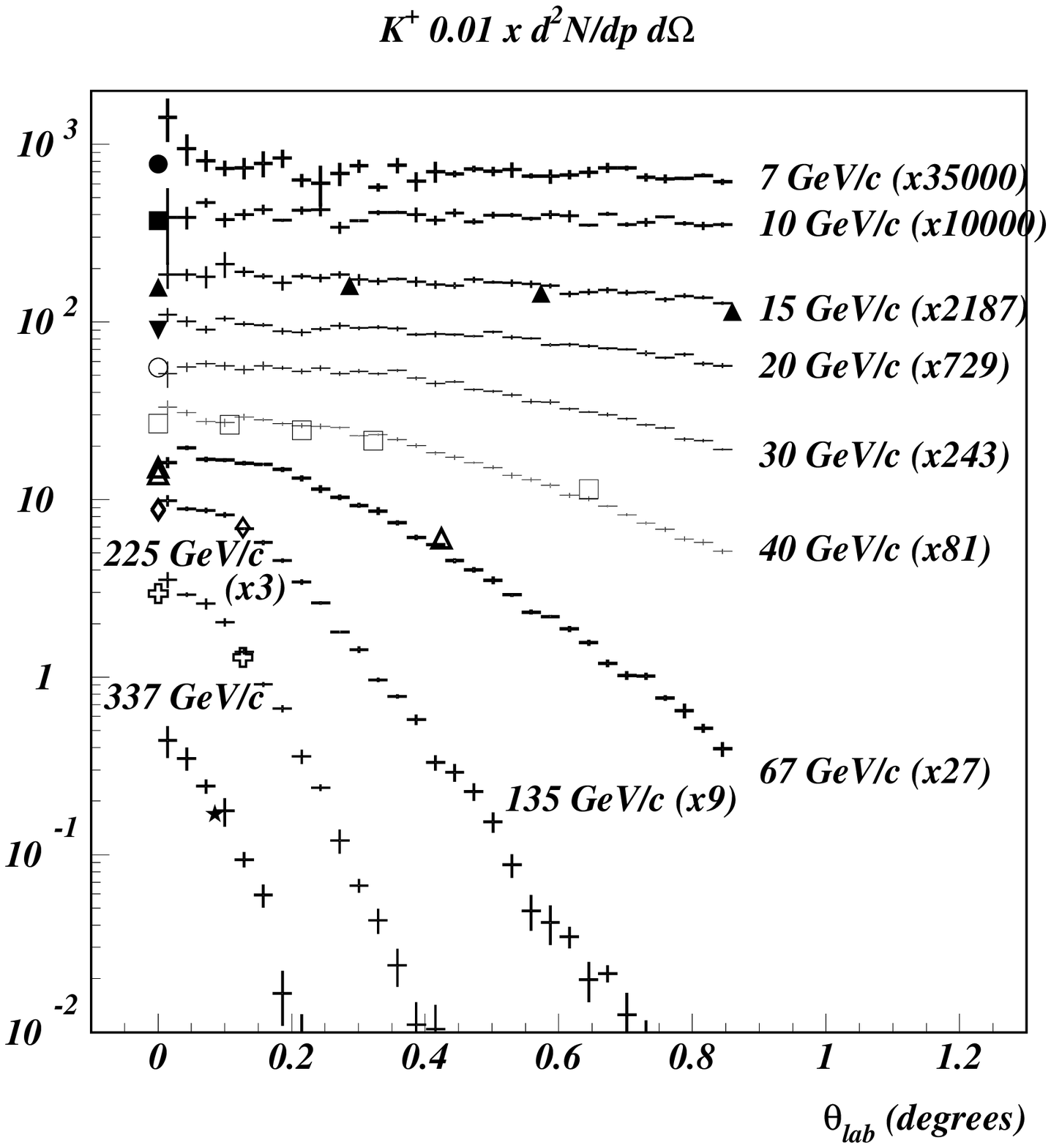,width=75mm}} &
\makebox[75mm]{\epsfig{file=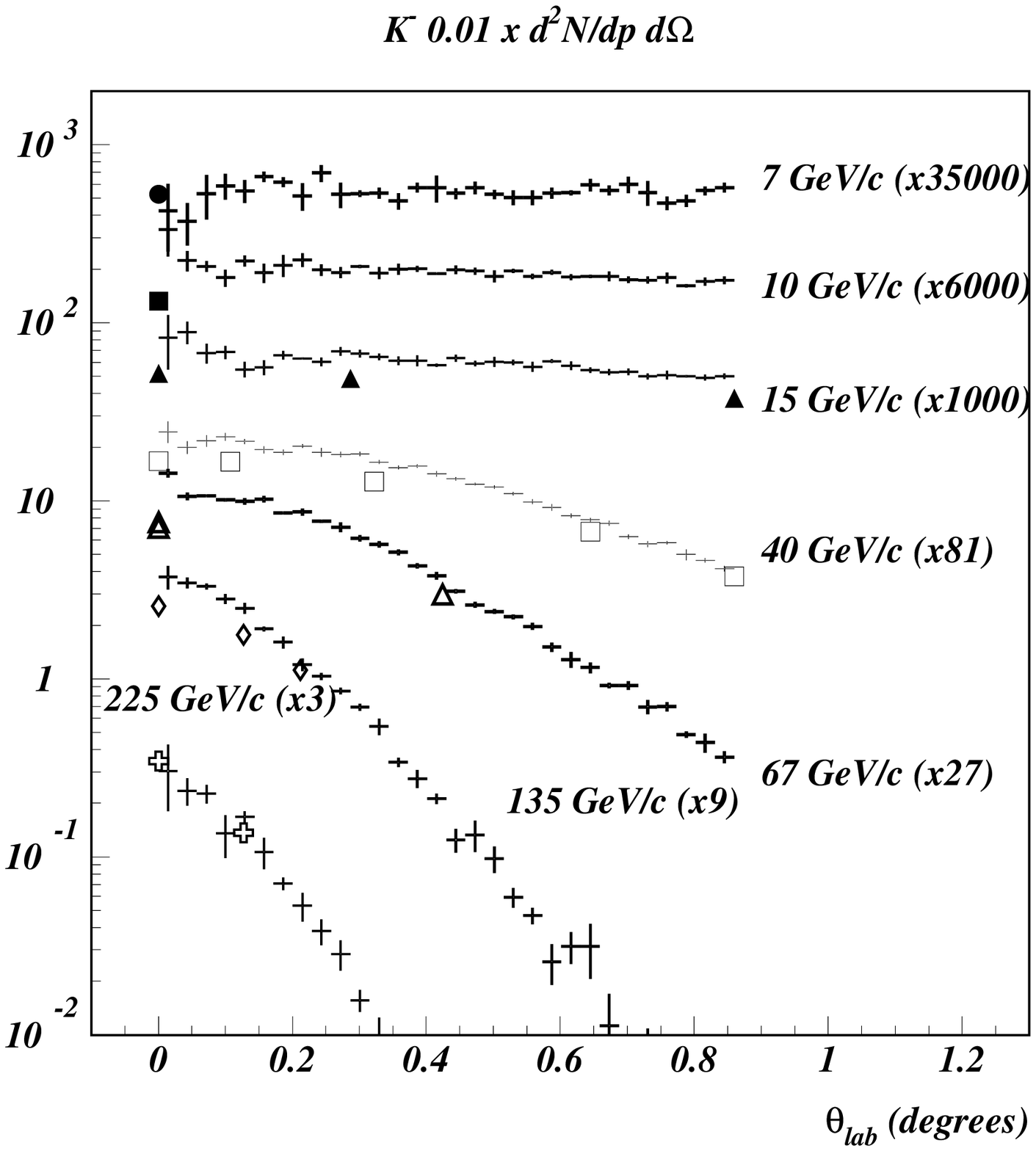,width=75mm}} \\
\end{tabular}
\caption{
Double differential cross section for 
$K^+$ (left) and $K^-$ (right)
production for 450~GeV/c protons on a 10 cm thick Be target 
(data from ref.~\protect\cite{atherton} and \protect\cite{spy}).
Histograms are simulation results.
 Data are given as a function
of $\theta_{lab}$ for different momentum bins. The
distributions are scaled by increasing factors (reported in the
figure) to allow a clearer separation.
\label{fig:spyk}}
\end{center}
\end{figure}

Two sets of data are of particular relevance to check the quality
of a model to be used for the calculation of atmospheric neutrino
fluxes.  These concern p-Be collisions and are reported in
Fig.~\ref{fig:pbey} and Fig.~\ref{fig:eich}: in ref.\cite{abbott} the
central rapidity region has been 
mainly explored, while
in ref.\cite{eichten} the forward region has been investigated. 
In both cases the agreement of \FLUKA{} predictions is
quite good.
 Measurements of  $\pi^{\pm}$ and $K^{\pm}$ production rates by 400 GeV/c
 protons on Be targets were performed by Atherton et al. \cite{atherton}
 for secondary particle momenta above 60 GeV/c and up to 500 MeV/c of
 transverse momentum.
 Recently the NA56/SPY (Secondary Particle Yields) experiment  \cite{spy} 
 was  devoted to directly measure these yields in the momentum
 region below 60 GeV/c. The SPY experiment measured
 the production at different angles $\theta$ and momenta $P \leq 135$ GeV/c
 down  to 7 GeV/c for pions, kaons, protons and their antiparticles,
 using a 450 GeV/c proton beam impinging on Be targets.
 These data were extremely valuable to improve the hadronization model
 of \FLUKA{} so to arrive at the present version.
 \FLUKA{} is in agreement with the Atherton 
%TM
{\sl et al} 
and the SPY  measurements
 at the level of $\sim 20 \%$  in the whole momentum range of all secondaries,
 with the exception of a few points mostly for negative kaons.
 The case of pions is reported in Fig.~\ref{fig:spypi}, and the analogous
plot for kaons is given in Fig.~\ref{fig:spyk}.
 Also the $\theta$ dependence of the measured yields  
 is reasonably described by \FLUKA{}.
 The measured $K^{\pm}/\pi^{\pm}$ ratios are reproduced to
 better than $20 \%$ below 120 GeV/c.

\section{Features of nucleon-Air collisions in the FLUKA model}

In the final state of one interaction one finds the fragments of the target
nucleus, a leading nucleon that carries the baryon number of the
projectile, a number of mesons, mostly pions, with a non negligible
contribution of kaons. In addition one has to consider the production of
nucleon antinucleon pairs and of heavy flavors, these effects are
however of small importance at the energies that are relevant for
atmospheric neutrinos. Most neutrinos are produced in the chain decay of
the charged pions such as $\pi^+ \rightarrow \mu^+ + \nu_\mu$ followed by
$\mu^+ \rightarrow e^+ \nu_e \nubar_\mu$ with $K^\pm$ and 
$K_L$ giving smaller contributions.
In order to characterize the features of nucleon-Air collisions that are
relevant in this discussion, we can identify few
fundamental quantities: the nucleon-Air cross sections, 
the distribution of the energy
fraction carried away by produced particles 
and the ``spectrum weighted moments'' $Z_{ij}$, all
as a function of projectile energy.

\begin{figure}[htbp]
\begin{center}
\begin{tabular}{ccc}
\makebox[50mm]{\epsfig{file=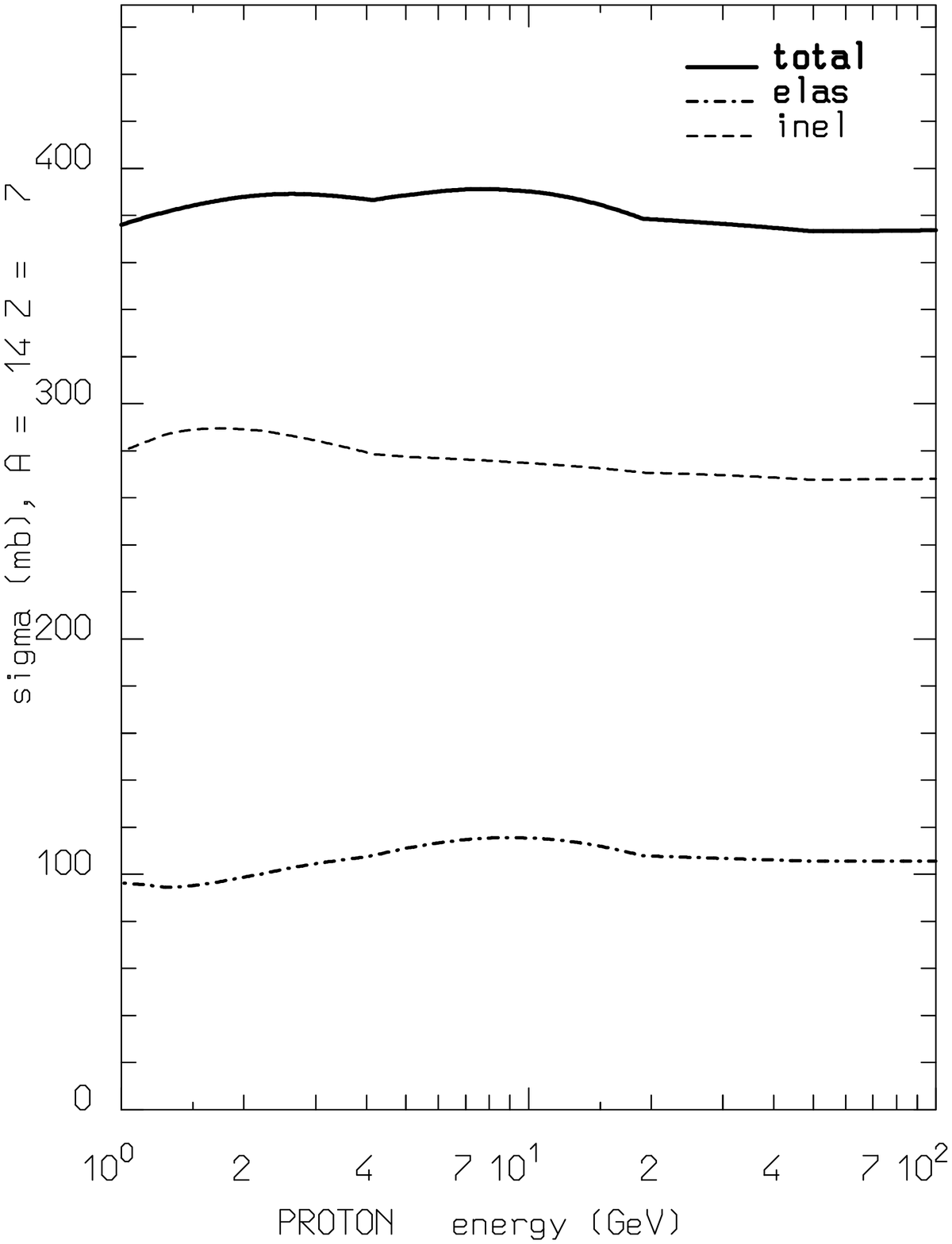,width=5cm}} &
\makebox[50mm]{\epsfig{file=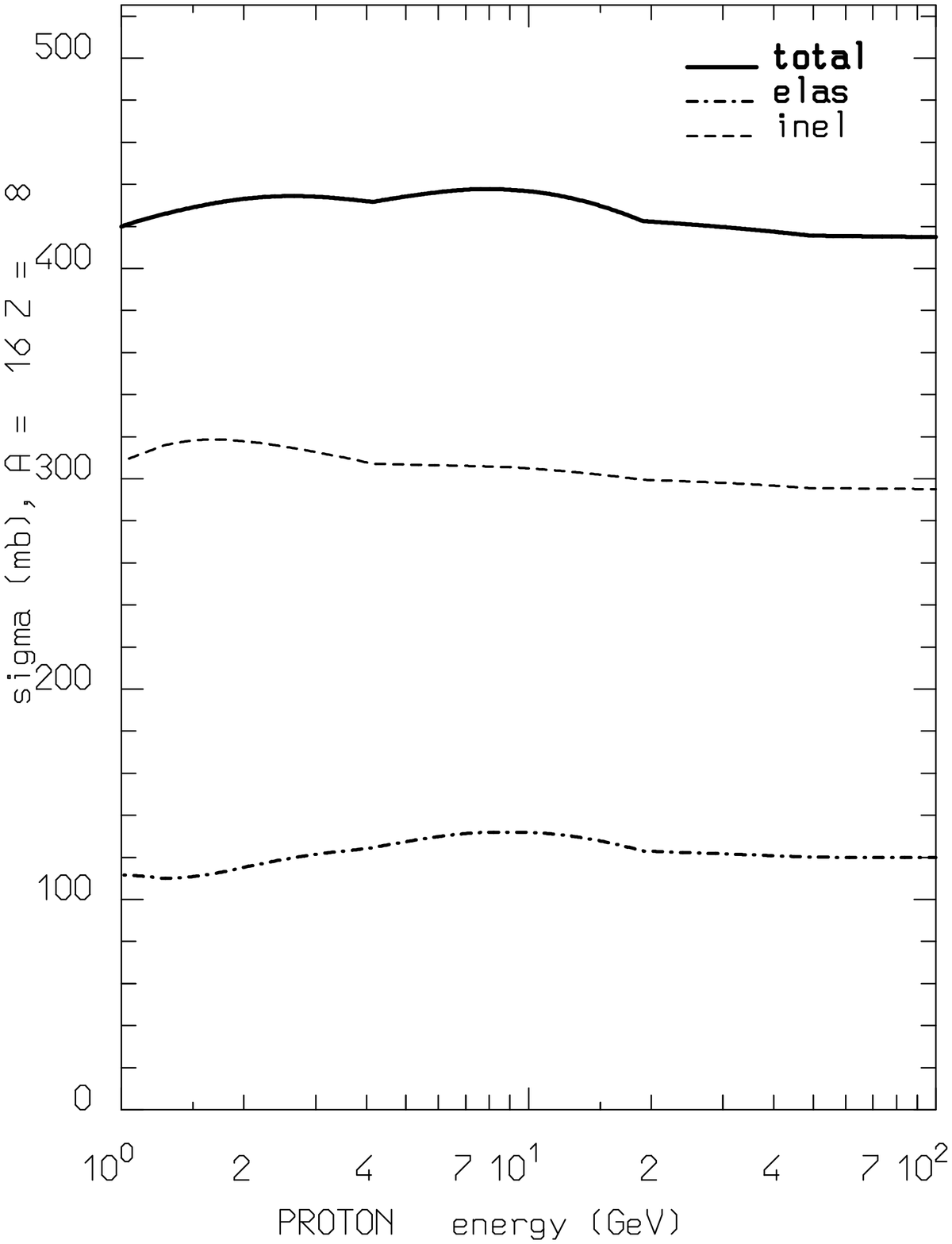,width=5cm}} &
\makebox[50mm]{\epsfig{file=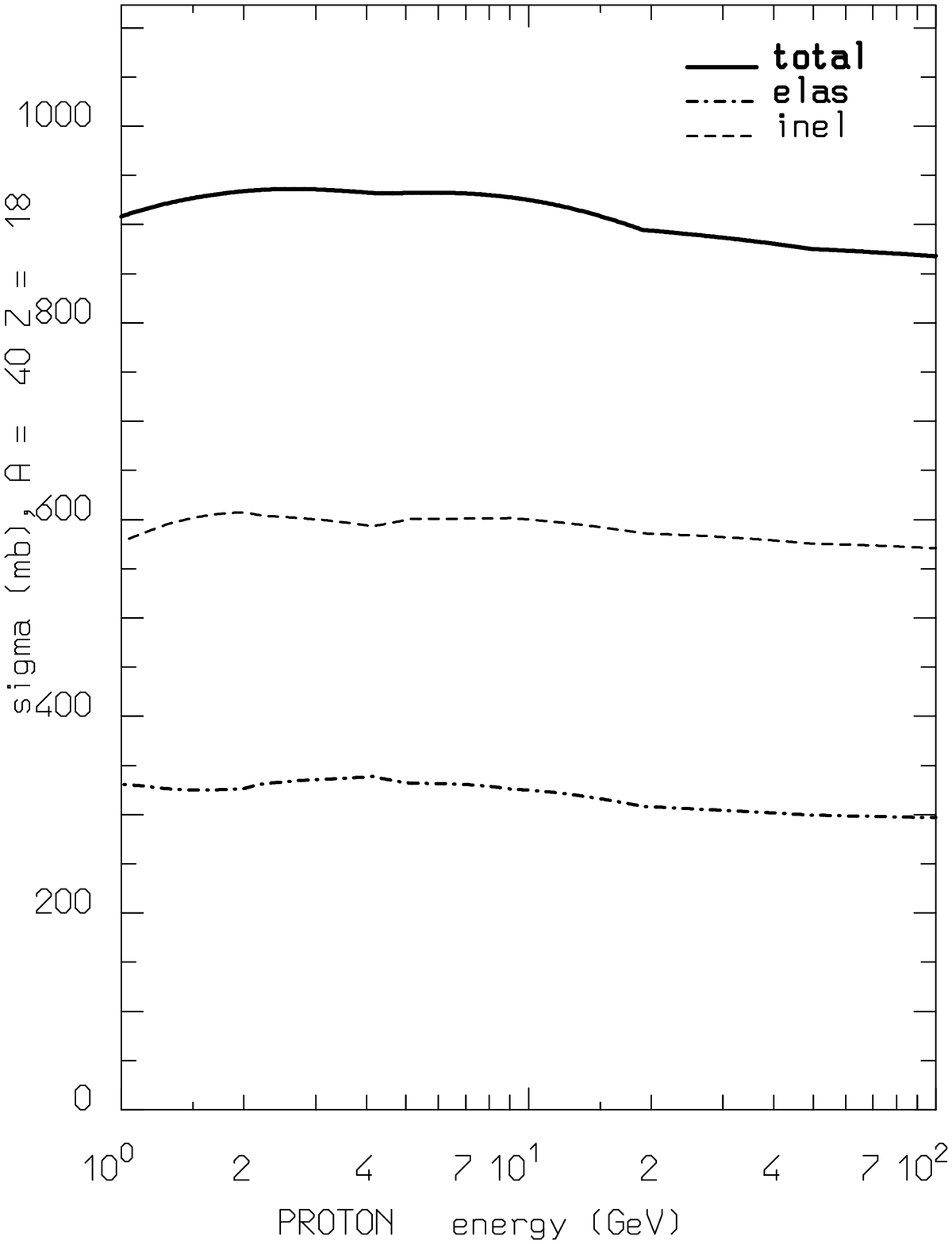,width=5cm}} \\
\end{tabular}
\caption{FLUKA cross section as a function
of proton kinetic energy for collisions with
Nitrogen (left), Oxygen (center) and Argon (right),
in the kinetic energy range 1--100 GeV. 
Top curve is the total cross section, intermediate
curve is inelastic cross section while the bottom curve
is the elastic one,
\label{fig:cross1}}
\end{center}
\end{figure}

In FLUKA, total, elastic and inelastic cross sections
for hadron-Nucleus collisions are derived in the framework of the Glauber
formalism, using as input the
tabulated data (when they exist) of Particle Data Group for hadron-hadron
collisions.
In Fig.~\ref{fig:cross1} we show the resulting proton-Air cross section
for Nitrogen, Oxygen and Argon as a function
of proton kinetic energy in the range 1-100 GeV (cross section
for neutrons are very similar). We plot separately the
elastic, inelastic and total contributions to cross section.
Here what we call ``inelastic'' cross section is meant to include
also what is usually called the ``quasi-elastic'' part, {\it i.e.}
the one which corresponds to nuclear excitation without 
particle production.

The resulting interaction length in Air
increases slowly and continuously from 85.1 g/cm$^2$ at
about 5 GeV to 88.2 g/cm$^2$ at 100 GeV.
Below 5 GeV the behavior is non monotonic due to
the resonances effect, and the values varies
between 81.9 and 85.1 g/cm$^2$.

\begin{figure}[htbp]
\begin{center}
\makebox[150mm]{\epsfig{file=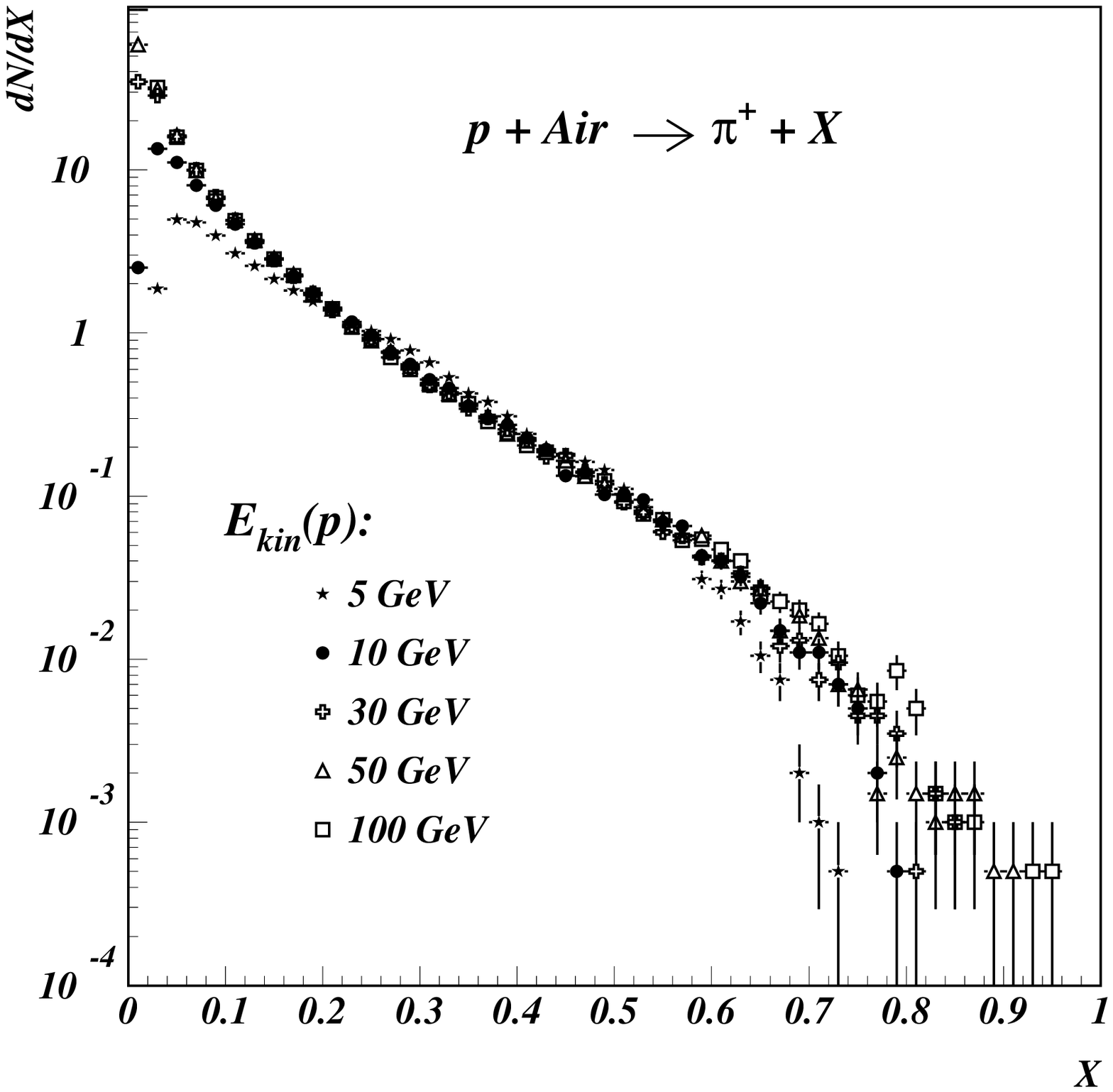,width=150mm%
%,bbllx=60pt,bblly=100pt,bburx=590pt,bbury=740pt
,height = 15.cm}}
\caption{ $dN/dx_{lab}$ for positive pions produced
by FLUKA in p--Air collisions at different kinetic
energies of the proton projectile.
\label{fig:xpip}
}
\end{center}
\end{figure}

\begin{figure}[htbp]
\begin{center}
\makebox[150mm]{\epsfig{file=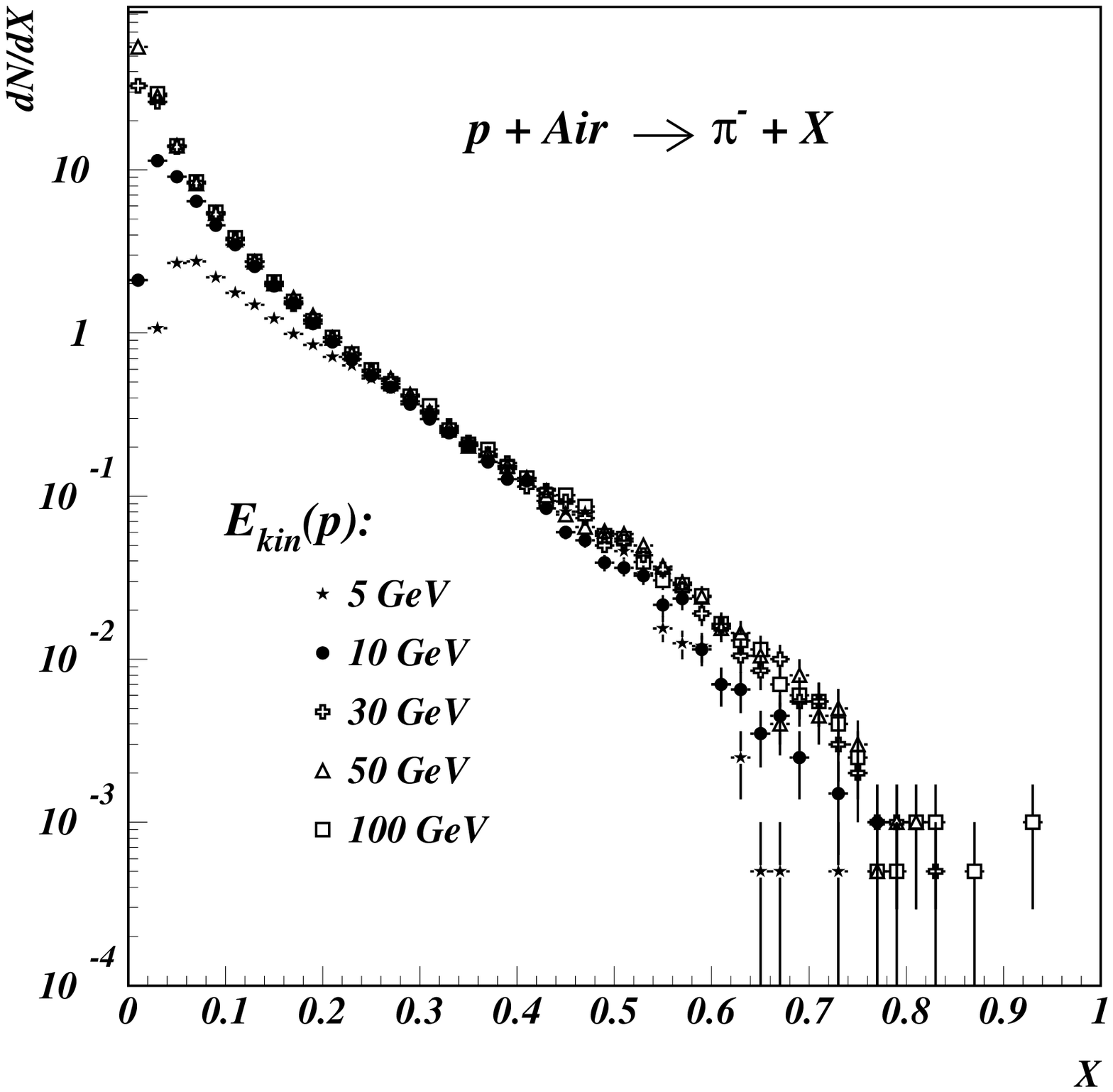,width=150mm%
%,bbllx=60pt,bblly=100pt,bburx=590pt,bbury=740pt
,height = 15.cm}}
\caption{ $dN/dx_{lab}$ for negative pions produced
by FLUKA in p--Air collisions at different kinetic
energies of the proton projectile.
\label{fig:xpim}
}
\end{center}
\end{figure}

\begin{figure}[htbp]
\begin{center}
\makebox[150mm]{\epsfig{file=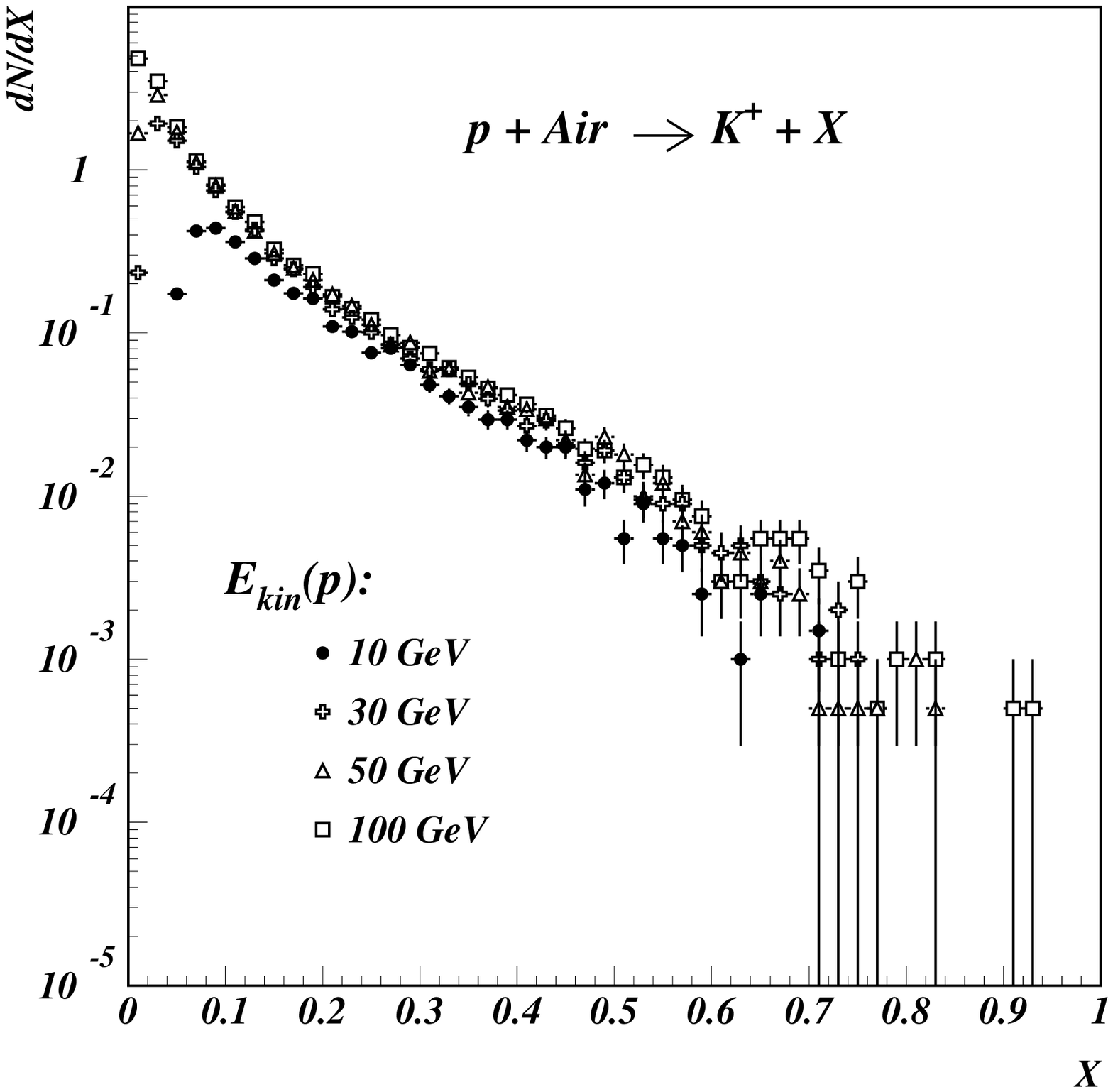,width=150mm%
%,bbllx=60pt,bblly=100pt,bburx=590pt,bbury=740pt
,height = 15.cm}}
\caption{ $dN/dx_{lab}$ for positive Kaons produced
by FLUKA in p--Air collisions at different kinetic
energies of the proton projectile.
\label{fig:xkp}
}
\end{center}
\end{figure}

\begin{figure}[htbp]
\begin{center}
\makebox[150mm]{\epsfig{file=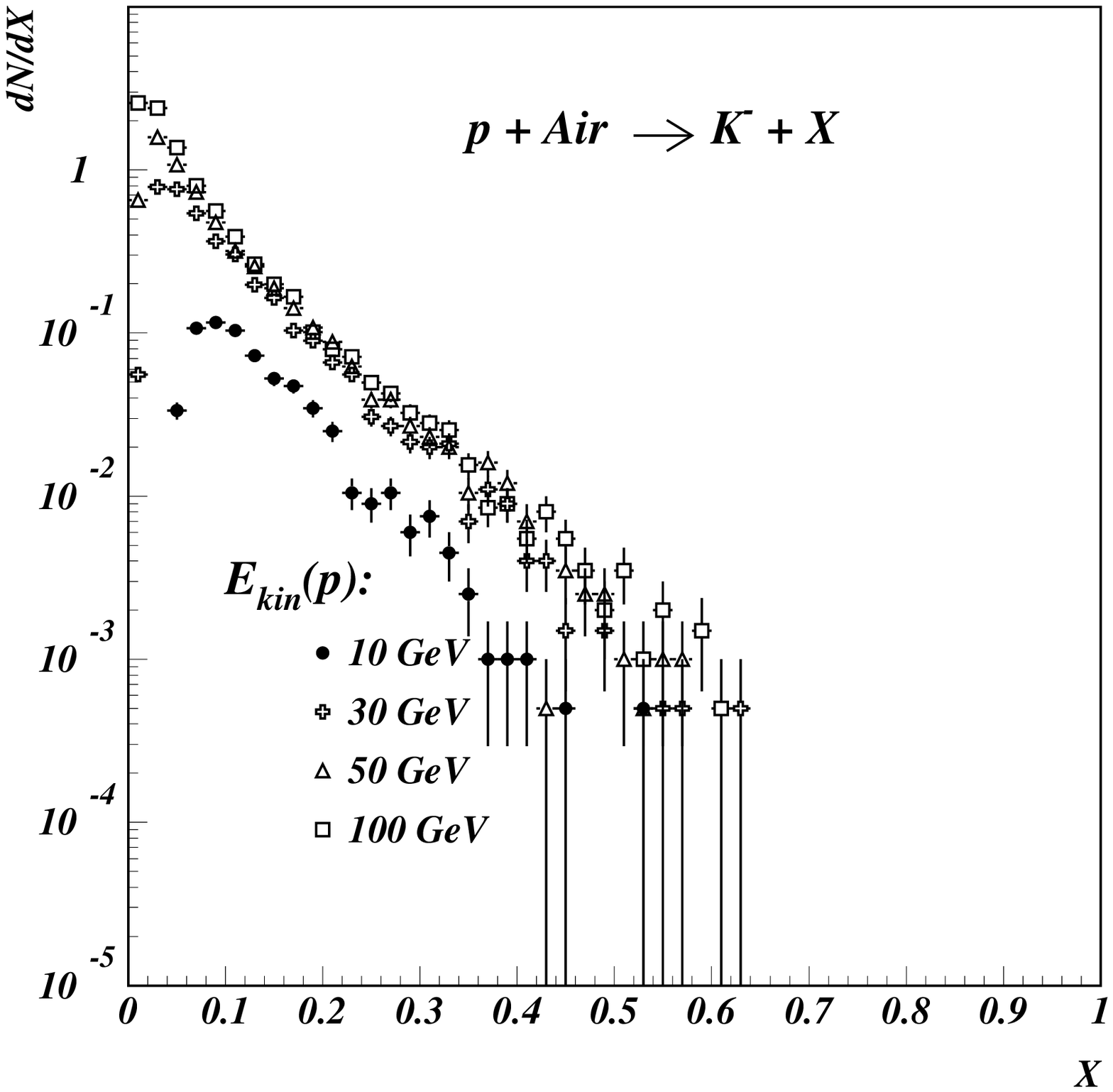,width=150mm%
%,bbllx=60pt,bblly=100pt,bburx=590pt,bbury=740pt
,height = 15.cm}}
\caption{ $dN/dx_{lab}$ for negative Kaons produced
by FLUKA in p--Air collisions at different kinetic
energies of the proton projectile.
\label{fig:xkm}
}
\end{center}
\end{figure}

\begin{figure}[htbp]
\begin{center}
\makebox[150mm]{\epsfig{file=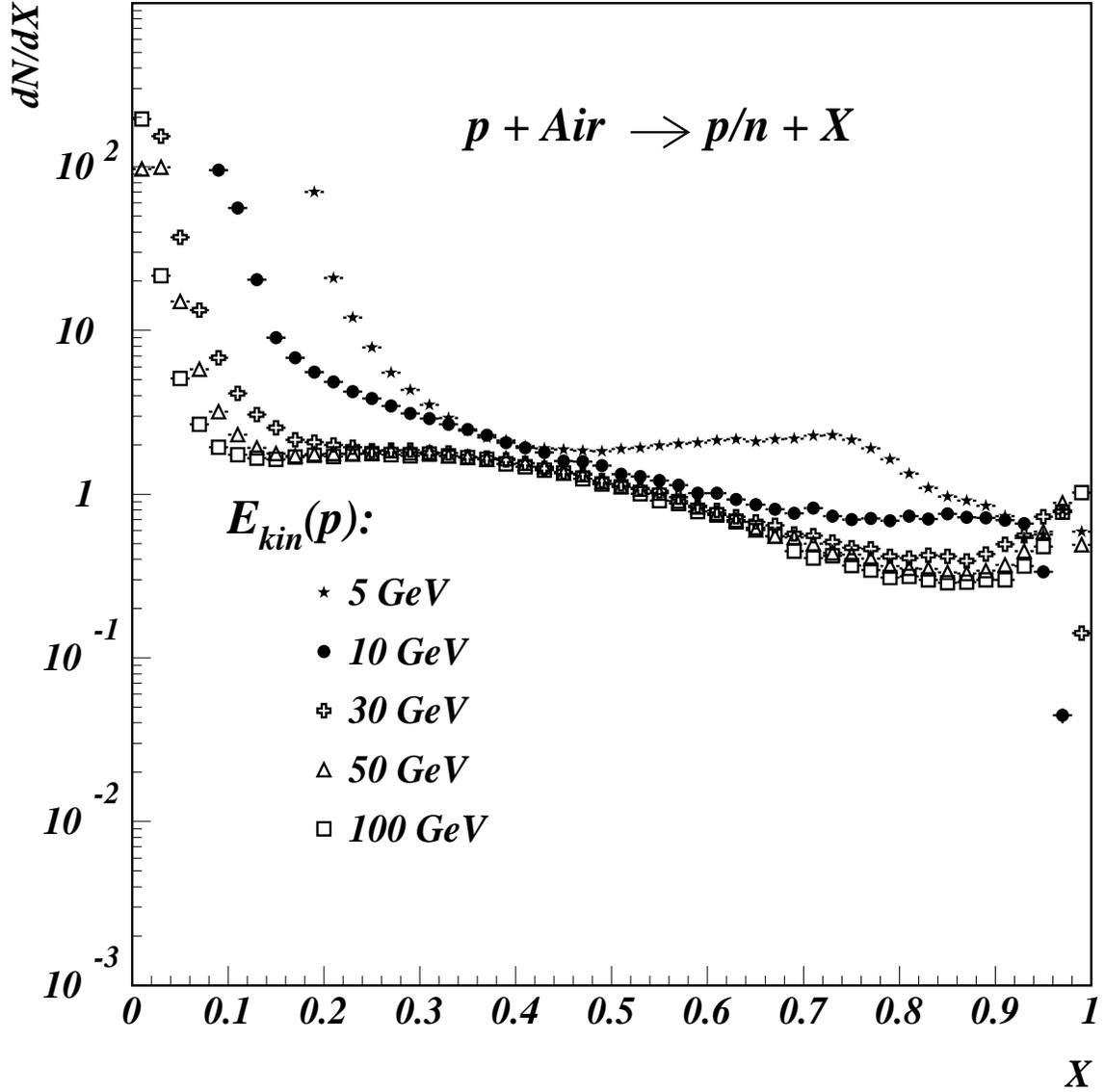,width=150mm%
%,bbllx=60pt,bblly=100pt,bburx=590pt,bbury=740pt
,height = 15.cm}}
\caption{ $dN/dx_{lab}$ for nucleons produced
by FLUKA in p--Air collisions at different kinetic
energies of the proton projectile.
\label{fig:xnucl}
}
\end{center}
\end{figure}

For our purposes, particle production can be conveniently studied in terms of
a longitudinal non dimensional variable like
$x_{lab} = E_j/E_i$, which is the ratio of the
secondary particle $j$ total energy over the primary particle $i$ total energy
($x_{lab}$ is approximately equal to $x_{F}$ in the forward region). 
We can then construct the $dN_{ij}/dx_{lab}$ distributions. These are the
differential multiplicity distributions of secondary $j$ as produced by primary $i$ in
collisions with Air nuclei as a function of $x_{lab}$. 
Examples of the $x_{lab}$ distribution in p--Air collisions
predicted by FLUKA are shown in Fig.~\ref{fig:xpip}, \ref{fig:xpim},
\ref{fig:xkp}, \ref{fig:xkm} and \ref{fig:xnucl} respectively
for produced $\pi^+$, $\pi^-$, $K^+$, $K^-$ and nucleons,
at different kinetic energy of the projectile. Here $x_{lab}$ is defined 
as the ratio of secondary total energy with respect to the primary total
energy.

In table \ref{tab:efr} we report the average energy fractions for
the secondary particles, produced in proton-Air collisions,
which mostly contribute to neutrino production,
between 5 and 100 GeV of kinetic energy of the projectile in
the laboratory system. 
Energy fraction is defined as:
\begin{equation}
F_{ij} = \int_{0}^{1} x_{lab} \frac{dN_{ij}}{dx_{lab}} dx_{lab}
\end{equation}
We also include $\Lambda$'s since they decay in a very short time,
giving an effective contribution to pion production.

\begin{table}
\begin{center}
\begin{tabular}{|c||c|c|c|c|c|c|c|c|c|}
\hline
$E^p_k$ (GeV)& $\pi^+$ & $\pi^-$ & $\pi^0$ & $K^+$ & $K^-$ & $K^0$+$\bar
K^0$ & $p$ & $n$ & $\Lambda$ \\
\hline
5  &0.149  &0.092  &0.150  &0.0039  &0.0001  &0.0030  &0.359  &0.215  &0.0066  \\
10 &0.157  &0.112  &0.163  &0.0090  &0.0014  &0.0101  &0.322  &0.193  &0.0141  \\
30 &0.161  &0.124  &0.172  &0.0150  &0.0056  &0.0192  &0.291  &0.167  &0.0179  \\
50 &0.162  &0.128  &0.175  &0.0169  &0.0079  &0.0237  &0.282  &0.155  &0.0185  \\
100&0.166  &0.134  &0.180  &0.0194  &0.0116  &0.0296  &0.263  &0.141  &0.0181  \\ 
\hline
\end{tabular}
\caption{Average energy fractions of secondary particles produced
in p--Air collisions as a function of the projectile kinetic energy.
\label{tab:efr}
}
\end{center}
\end{table}

The numbers for neutron-Air collision are closely related, and, 
for the case of an isoscalar target, charge symmetry is substantially
valid. Therefore, if we consider the production of 
a meson with isospin $|I,I_z>$, 
$<E^{p+Air}(I_z)>~\simeq <E^{n+Air}(-I_z)>$, so that,
for example, $<E^{p+Air}(\pi^+)>~\simeq <E^{n+Air}(\pi^-)>$.
This is relevant when nuclear primaries (even following the superposition
principle) are considered, since their interaction is  eventually reduced to
Z proton--Air and A-Z neutron--Air collisions.
We notice that these average energy fractions are approximately constant 
in this energy range, reflecting an 
approximate validity of Feynman scaling in the model.
At low $E_{kin}$ (few GeV's), only pions are significantly produced, but 
their energy fraction decreases: this is a reflection of the fact that
an approximately constant amount of energy $\Delta E_{frag}$ is absorbed in the
excitation of the target nucleus, and with decreasing projectile energy
this represents a larger and larger contribution to the average energy
fraction.

The spectrum weighted moments are defined as:
\begin{equation}
Z_{ij} = \int_{0}^{1} (x_{lab})^{\gamma - 1} \frac{dN_{ij}}{dx_{lab}} dx_{lab}
\end{equation}
where $\gamma$ = 2.7 is the approximate spectral index of the differential 
cosmic ray spectrum.
Their use in the literature is justified by the fact that
the inclusive yield of secondary cosmic ray 
particles at a given energy is almost proportional to $Z$, when the
primary spectrum is a power law with a constant spectral index
in the whole useful energy spectrum. This is a good approximation
for nucleon energies exceeding the TeV scale, but it is not the case
in the range of energies considered in this work, since a single power law
does not fit the primary spectrum. However, they remain a useful and
commonly accepted tool to characterize and compare different interaction 
models (notice that for $\gamma$=2 they reduce to the energy fractions).
For instance, we can state that, in a first approximation, the flux
of $\numu + \anumu$ from pions only (excluding for instance
muon decay) can be considered as:
\begin{equation}
\phi_{\numu + \anumu} \propto \frac{ Z_{p \pi^+} + Z_{p \pi^-}}{ 1-Z_{pp}-Z_{pn} }
\end{equation}

The resulting spectrum  weighted moments are tabulated
in Table \ref{tab:zmom}.

\begin{table}
\begin{center}
\begin{tabular}{|c||c|c|c|c|c|}
\hline
$E^p_k$ (GeV) & $\pi^+$ & $\pi^-$ & $K^+$ & $K^-$ & $p$+$n$ \\
\hline
5  &0.0388 &0.0229 &0.0012 &0.0000 &0.374    \\
10 &0.0404 &0.0271 &0.0026 &0.0003 &0.330    \\
30 &0.0415 &0.0306 &0.0042 &0.0015 &0.294    \\
50 &0.0422 &0.0315 &0.0048 &0.0022 &0.282    \\
100&0.0426 &0.0327 &0.0052 &0.0030 &0.261    \\ 
\hline
\end{tabular}                        
\caption{Spectrum weighted moments (for a spectral index of the
differential primary spectrum $\gamma$= 2.7) 
for secondary particles produced
in p--Air collisions as a function of the projectile kinetic energy.
\label{tab:zmom}
}
\end{center}
\end{table}

\section{Example of calculations of particles in atmosphere}
\label{sec:example}

Beyond the topic of atmospheric neutrino fluxes, 
the \FLUKA{} interaction model has been used also
to produce results on other secondary particles produced
in atmosphere by cosmic rays, which can be used as further cross check
of the validity of the model.
At least two remarkable results can be quoted:
\begin{enumerate}
\item The reproduction of the features of primary proton flux as a function
of geomagnetic latitude as measured by AMS\cite{AMSprb}, thus showing that 
the geomagnetic effects and the overall geometrical description of the
3--D setup are well under control\cite{zuccon}. 
In addition, the same work shows that
also the fluxes of secondary $e^+e^-$ measured at high altitude
are well
reproduced. Sub-cutoff $e^{\pm}$ are mainly (97$\%$) coming from decays 
of pions produced in the proton collisions with the atmospheric nuclei:
charged pions contribute through the $\pi-\mu-e$ chain, while $\pi^{0}$
through $\pi^{0}\rightarrow \gamma~\gamma$
with subsequent e.m.  showers. 
The relative contribution of charged pions to the sub-cutoff 
electrons (positron) fluxes at AMS altitude is found to be
$37\%$ ($47\%$), while the remaining $60\%$~($50\%$) appears to come from
$\pi^{0}$ production.  
This point deserves some considerations. The level of agreement in the
comparison of data and predictions for $e^{\pm}$ fluxes turns out to be an
important benchmark for 
the interaction model from the point of view of particle production in
atmosphere, since it is strictly linked to the meson
production (mostly pions at this energy). 

\item The good reproduction of the data on muons in atmosphere as measured
by the CAPRICE experiment\cite{caprice}, both at ground level and at different
floating altitudes: see ref.\cite{caprice_fluka}
for the relevant plots and numbers.
The agreement exhibited by the \FLUKA{} simulation for muons
of both charges gives confidence on
the predictions of FLUKA for the parent mesons of muons (mostly pions).
This work complements the previously mentioned studies oriented to the validation of
the model in terms of particle yields.
%\begin{figure}[htb]
%\begin{center}
%\begin{tabular}{cc}
%\mbox{\epsfig{file=figneg1.eps,height=6cm}} &
%\mbox{\epsfig{file=figneg2.eps,height=6cm}} \\
%\mbox{\epsfig{file=figpos1.eps,height=6cm}} &
%\mbox{\epsfig{file=figpos2.eps,height=6cm}} \\
%\end{tabular}
%\caption{
%Comparison between simulated (histogram) and detected
%(full symbols)
%negative (upper panels) and positive (lower panels) muon flux
%as a function of momentum for different 
%atmospheric depths. 
%Boxes  represent the statistical error in MonteCarlo results.
%\label{fig:negm}}
%\end{center}
%\end{figure}

The fluxes of atmospheric muons are strictly related to the neutrino ones,
because almost all $\nu$'s are produced either in association, with, or in the
decay of $\mu^\pm$. Therefore it is possible to conclude that, for that
choice of primary spectrum, the $\nu$ fluxes predicted by \FLUKA{} are probably
in the right range.
It the last years it has been remarked the noticeable agreement between the
original HKKM\cite{hkkm} and Bartol\cite{bartol}
calculations of the $\nu$ fluxes, despite they started from different estimates
of the primary flux and different hadronic interaction models.
Probably such an agreement is
not casual, but the result of the $\mu$ constraint, producing a sort of
error compensation.
%TM
If that is the case, the resulting flux could be biased by some
systematics.
\end{enumerate}

\begin{figure}[htbp]
\begin{center}
\makebox[150mm]{\epsfig{file=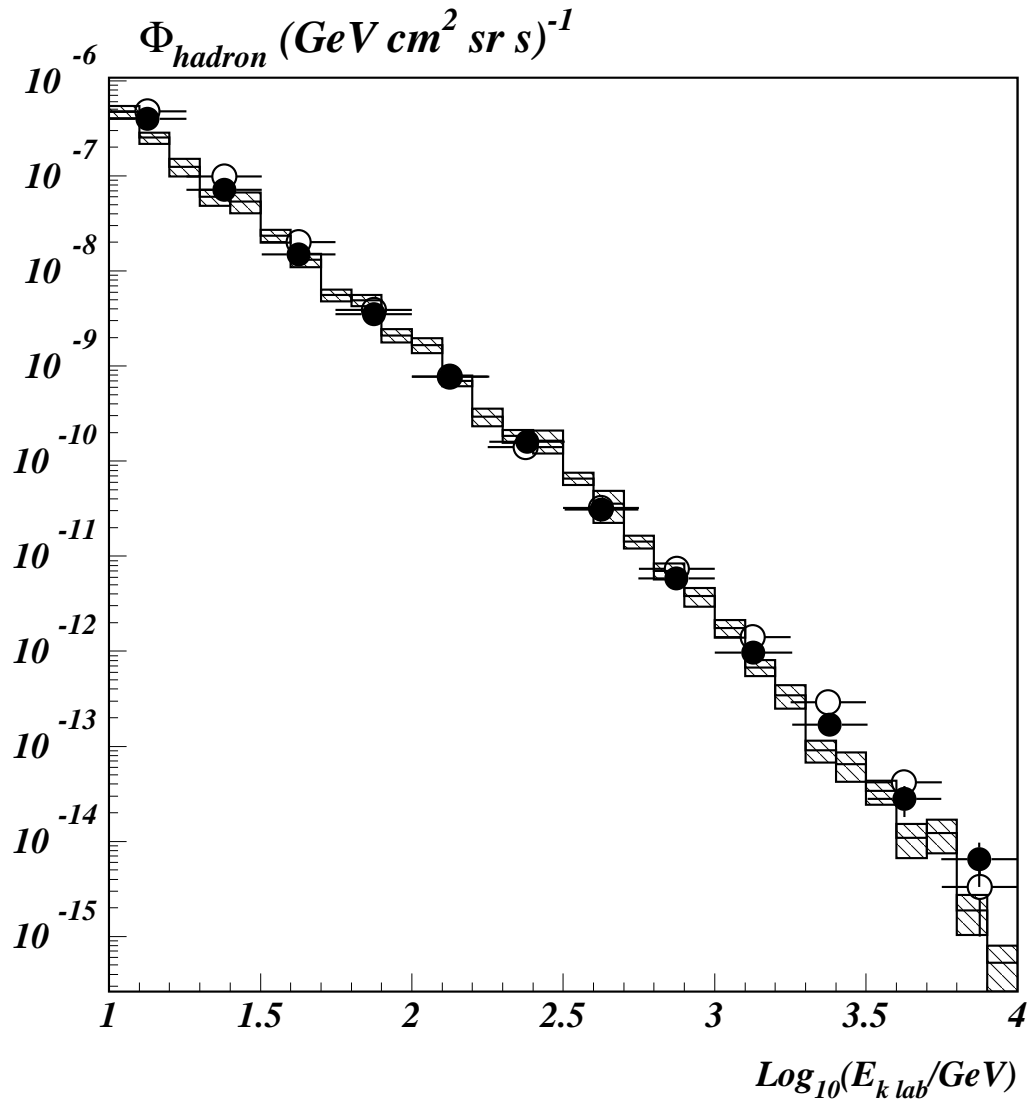,width=150mm%
,bbllx=60pt,bblly=100pt,bburx=590pt,bbury=740pt}}
\caption{Hadron flux referred to the vertical as measured with the
  calorimeter of the KASCADE 
experiment. Histogram is simulation result.
Two different sets of experimental data are shown, corresponding
to two different angular acceptances, from
ref.\protect\cite{KASCADE1} (black symbols: maximum zenith angle
30$^\circ$) and ref.\protect\cite{KASCADE2} 
(open symbols: maximum zenith angle 60$^\circ$).
\label{fig:atmhad}
}
\end{center}
\end{figure}

The shower simulations in atmosphere 
%TM
with FLUKA{} have been compared also 
to the most 
recent hadron spectra at different
latitudes and altitudes, obtaining remarkable agreement. As an example,
in fig.~\ref{fig:atmhad} we compare MonteCarlo results to the 
hadron flux referred to the vertical as 
measured with the calorimeter of the
KASCADE experiment\cite{KASCADE1,KASCADE2} with two
different angular acceptances.

\section{Simulation Results}
\label{sec:resul}

In Tables ~\ref{tabe}, \ref{tabae}, \ref{tabm} and \ref{tabam} we give the
differential $\nu_e$, $\nubar_e$, $\nu_\mu$ and $\nubar_\mu$ flux in units 
cm$^{-2}$ s$^{-1}$ sr$^{-1}$ GeV$^{-1}$ for some energy and cosine of the
zenith angle bins. 
The tables are given for Super-Kamiokande site and solar maximum, having
used the average primary cosmic ray fit in \cite{bartol_new}.
The full tables with a total number of 40 angular 
bins between $\cos\theta = -1$ to 1 and 68 energy bins between 0.1 and 200 GeV
are accessible in \cite{flukatab}, together with the
similar tables for Soudan and Gran Sasso cut-offs.

\begin{figure}[htbp]
\begin{center}
\makebox[150mm]{\epsfig{file=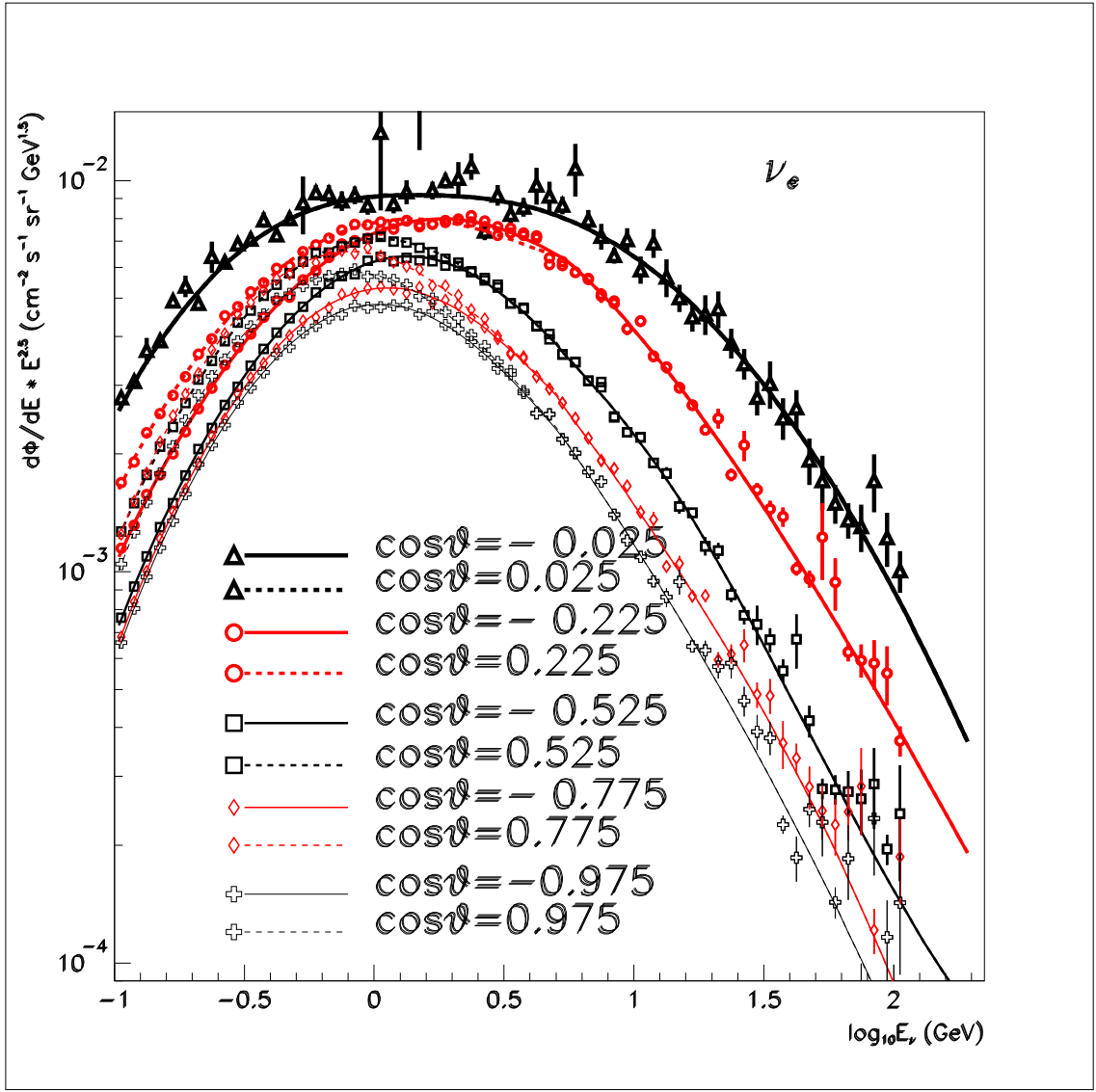,width=150mm%
%,bbllx=60pt,bblly=100pt,bburx=590pt,bbury=740pt
,height = 15.cm}}
\caption{Electron neutrino differential flux times E$^{2.5}$ 
versus neutrino energy at Super-Kamiokande site and for solar maximum. 
The average cosmic ray fit is used. Different curves are 
for different cosine of the zenith angle values. 
Lines are the result of the fits of the simulation results described 
in the text. Solid lines are for the lower hemisphere values, 
while dashed lines are for upper hemisphere
angles. Notice that for the same absolute value of $\cos\theta$
the solid and dashed lines do not overlap for energies $\lesssim 5$ GeV 
while they do overlap at larger energies (where the geomagnetic cut-off
does not produce any effect) and at the horizon.
Symbols show the simulation results in order to show the agreement with the
fits. Notice at very large angles and at large energies simulation errors
are noticeable and the fits are useful to smooth simulation points. 
\label{fluxe}
}
\end{center}
\end{figure}

\begin{figure}[htbp]
\begin{center}
\makebox[150mm]{\epsfig{file=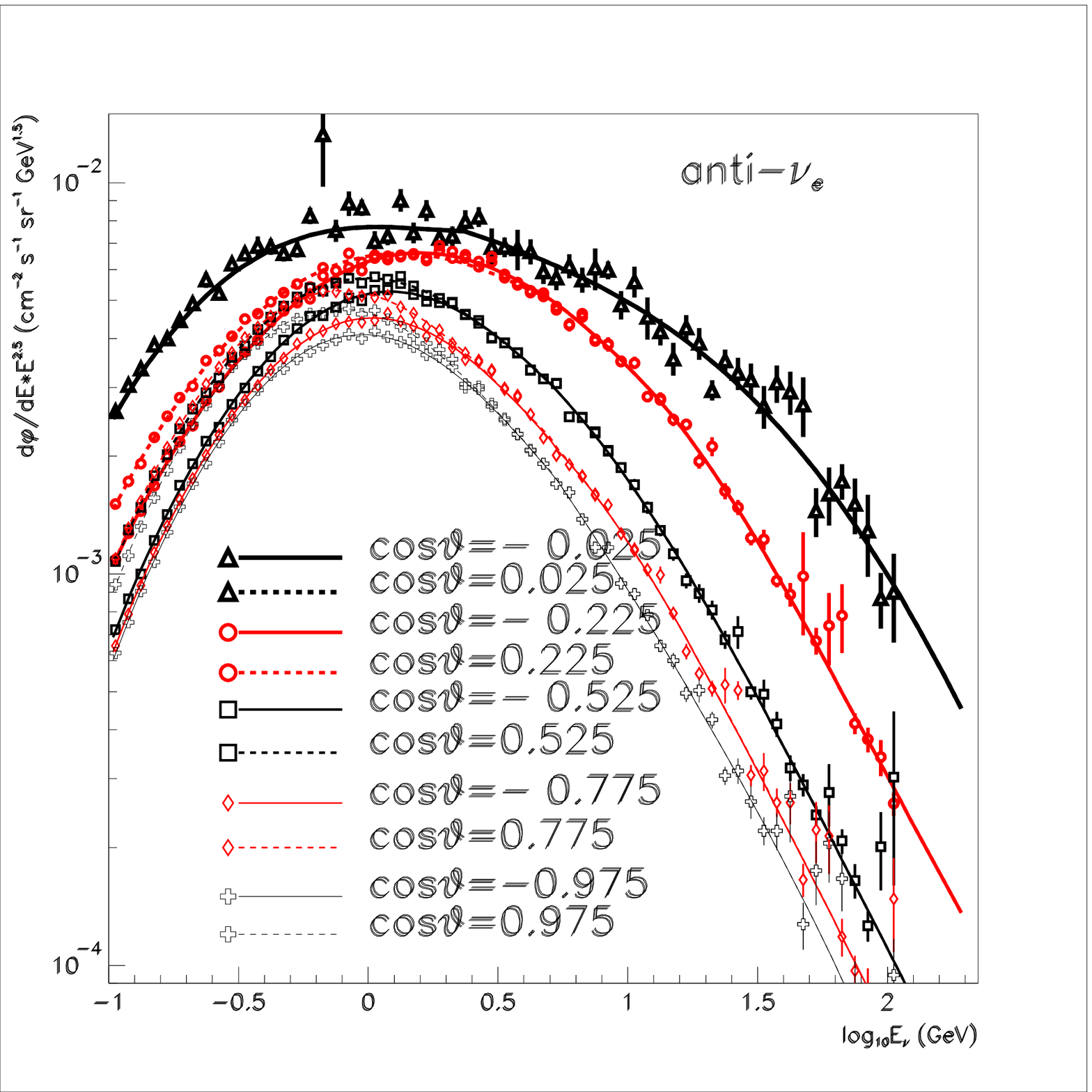,width=150mm%
%,bbllx=60pt,bblly=100pt,bburx=590pt,bbury=740pt
,height=15.cm}}
\caption{Electron anti-neutrino differential flux times E$^{2.5}$ 
versus neutrino energy at Super-Kamiokande site and for solar maximum. 
More details in the caption of Fig.~\protect\ref{fluxe}. 
\label{fluxae}
}
\end{center}
\end{figure}

\begin{figure}[htbp]
\begin{center}
\makebox[150mm]{\epsfig{file=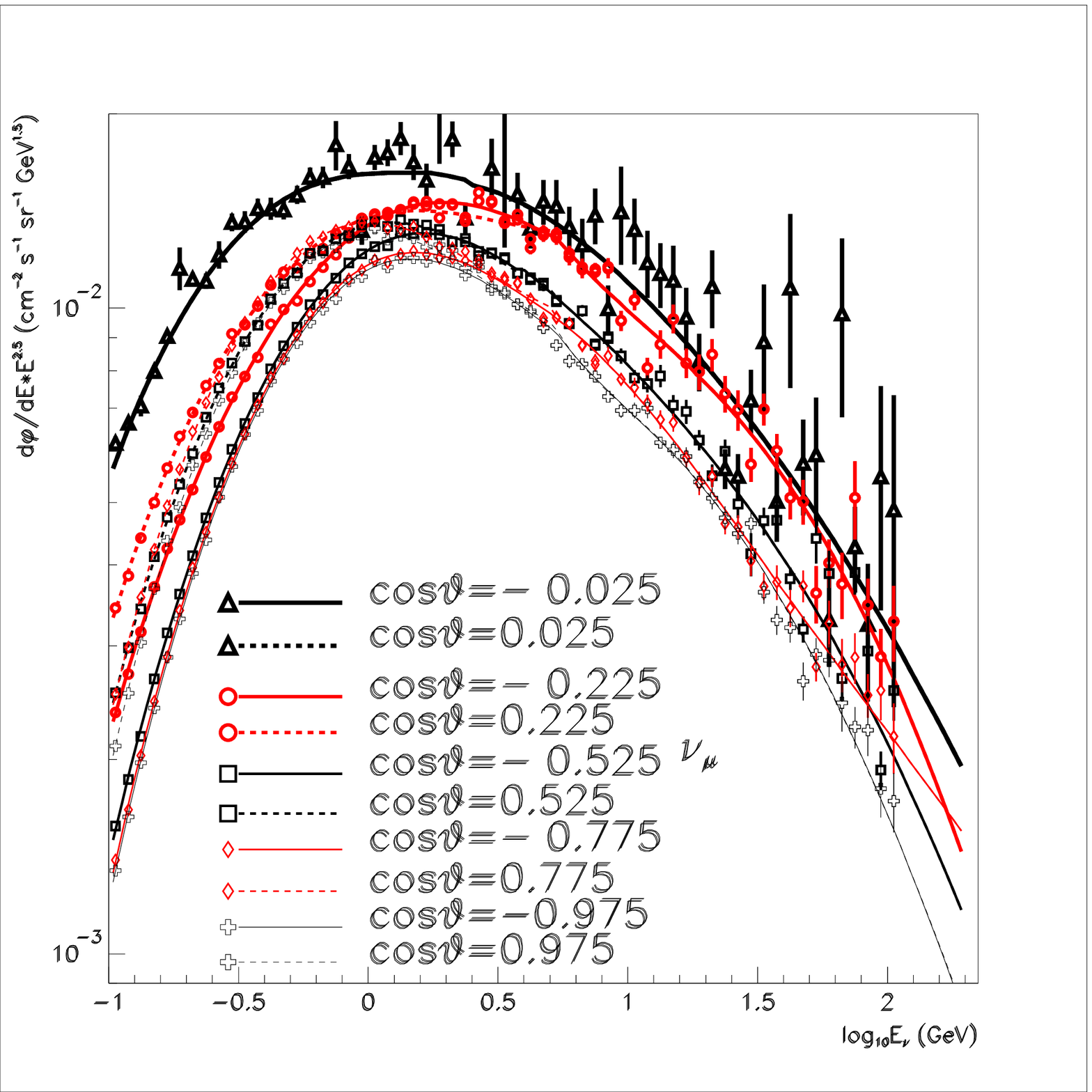,width=150mm%
%,bbllx=60pt,bblly=100pt,bburx=590pt,bbury=740pt
,height=15.cm}}
\caption{Muon neutrino differential flux times E$^{2.5}$ 
versus neutrino energy at Super-Kamiokande site and for solar maximum. 
More details in the caption of Fig.~\protect\ref{fluxe}. 
\label{fluxm}
}
\end{center}
\end{figure}

\begin{figure}[htbp]
\begin{center}
\makebox[150mm]{\epsfig{file=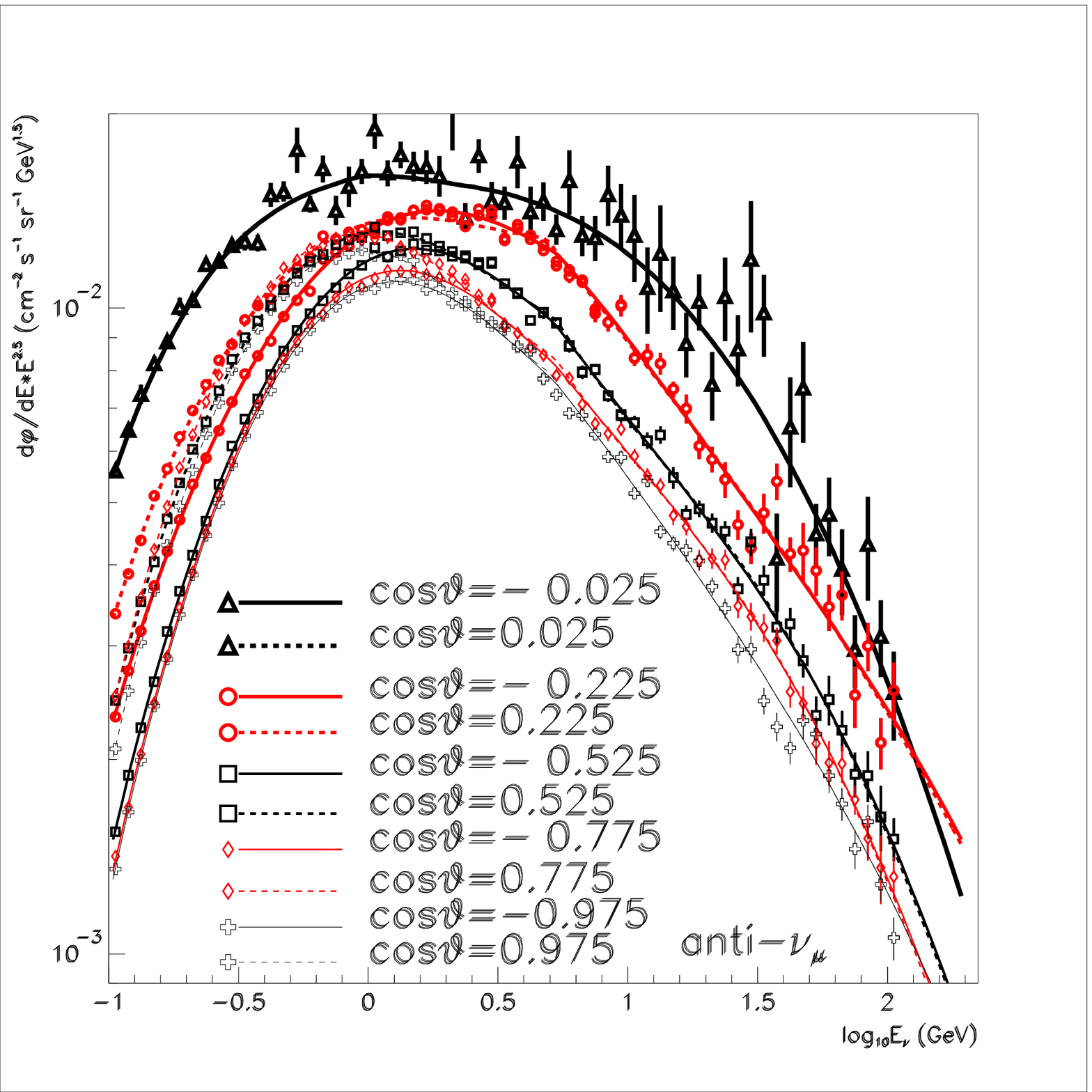,width=150mm%
%,bbllx=60pt,bblly=100pt,bburx=590pt,bbury=740pt}}
,height=15.cm}}
\caption{Muon anti-neutrino differential flux times E$^{2.5}$ 
versus neutrino energy at Super-Kamiokande site and for solar maximum. 
More details in the caption of Fig.~\protect\ref{fluxe}. 
\label{fluxam}
}
\end{center}
\end{figure}

In Fig.~\ref{fluxe}, \ref{fluxae}, \ref{fluxm} and \ref{fluxam} the
differential flux as a function of neutrino energy 
is shown for some values of $\cos\theta$ at Super-Kamiokande site and
solar maximum for the 4 flavors. 
A characteristic feature of the 3--D calculation is the excess
of events from near the horizontal at low energy.  This was
shown clearly in the first FLUKA paper\cite{flukanu}. 
Such a feature is clearly maintained in the plots reported here.
Additional explanations of how this effect originates are given
in \cite{lipgeo}.

\begin{figure}[htbp]
\begin{center}
\begin{tabular}{cc}
\makebox[75mm]{\epsfig{file=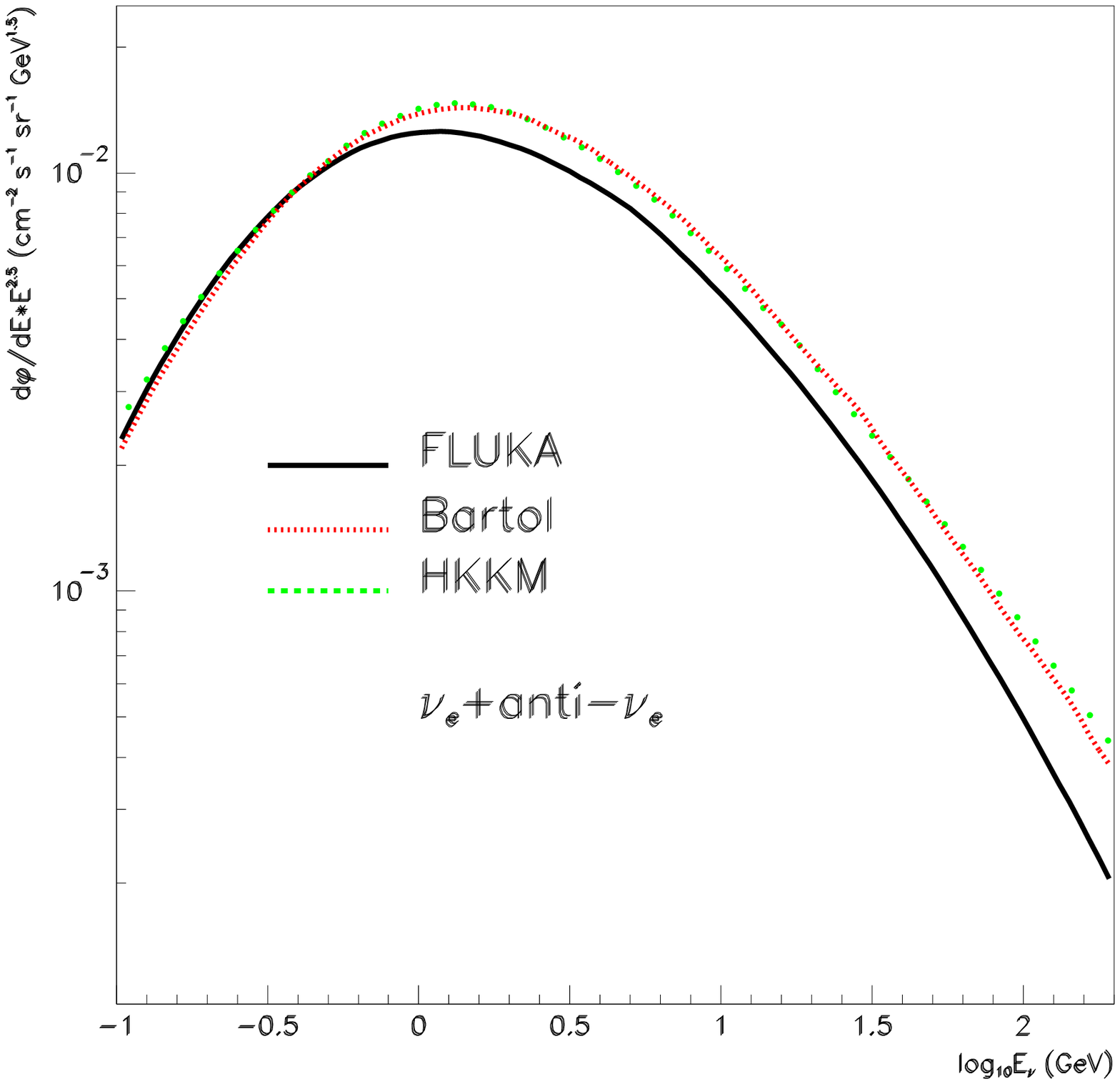,width=75mm}} &
\makebox[75mm]{\epsfig{file=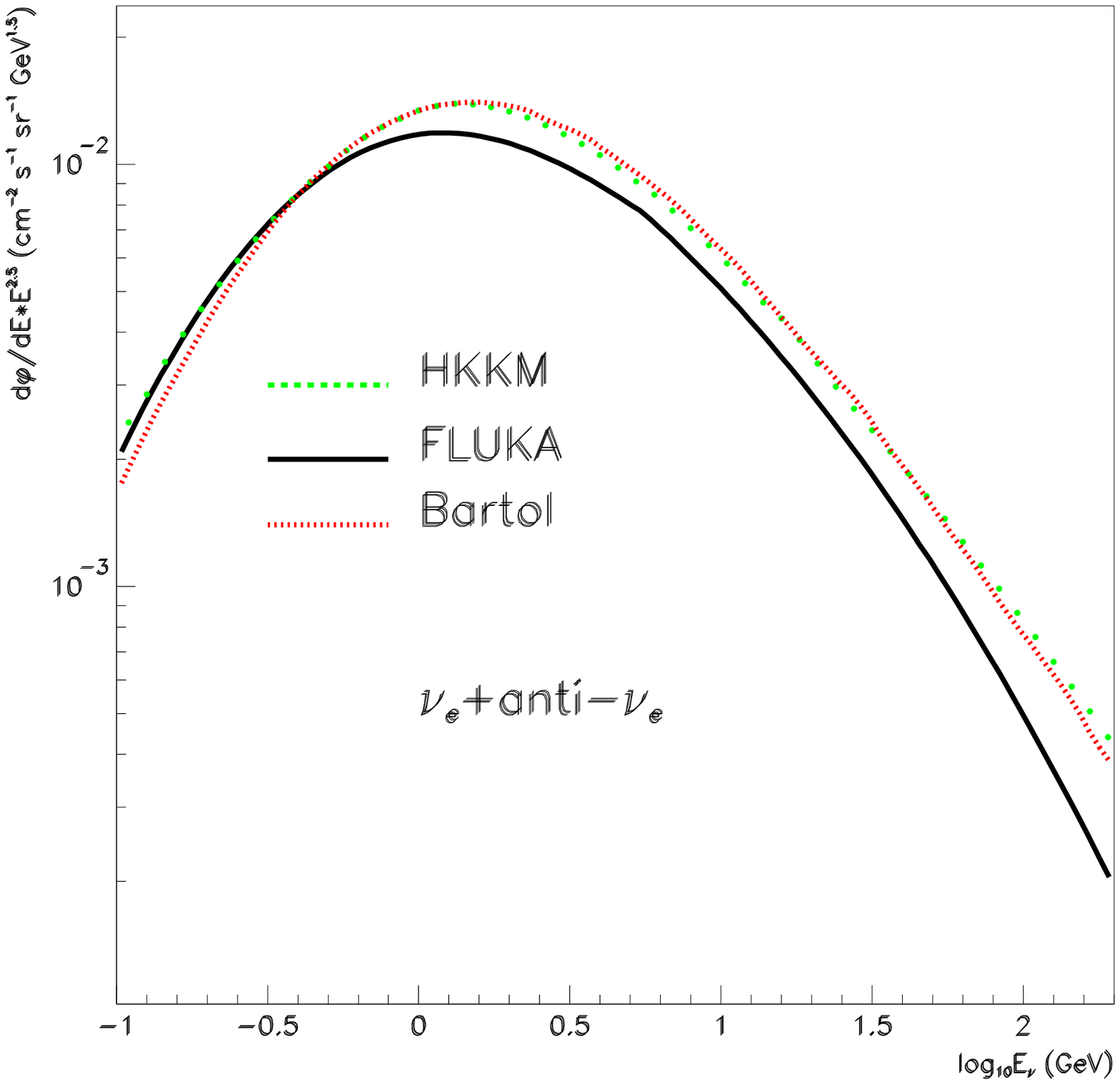,width=75mm}} \\
\end{tabular}
\caption{Electron neutrino and anti-neutrino differential flux times
E$^{2.5}$,
 angle averaged over the whole solid angle,
versus neutrino energy at Super-Kamiokande site for solar minimum (left)
and solar maximum (right)
for the FLUKA calculation (solid line), the Bartol one
(dotted line) \protect\cite{bartol} and for HKKM \protect\cite{hkkm}
(dashed line). 
\label{fluxconfe}
}
\end{center}
\end{figure}

Fig.~\ref{fluxconfe} shows the  angle averaged over the lower hemisphere
differential electron neutrino and anti-neutrino 
fluxes multiplied by E$^{2.5}$ at Super-Kamiokande for both solar minimum
and solar maximum modulation. 

In order to give a comparison with respect to the calculation referenced
so far by current experiments, the
FLUKA flux is shown together with the
Bartol group flux \cite{bartol} (dotted and red line) and the HKKM \cite{hkkm2}
(dashed and green line). It should be considered that, while the FLUKA flux
has been derived using the new fit to recent primary cosmic ray measurements,
the Bartol and HKKM fluxes were calculated using different evaluations of the
primary fluxes. It can be noticed that in our case the difference in flux
due to solar modulation is much more limited than for Bartol calculation.
Our minimum/maximum ratio is much closer to that of HKKM flux.

\begin{figure}[htbp]
\begin{center}
\begin{tabular}{cc}
\makebox[75mm]{\epsfig{file=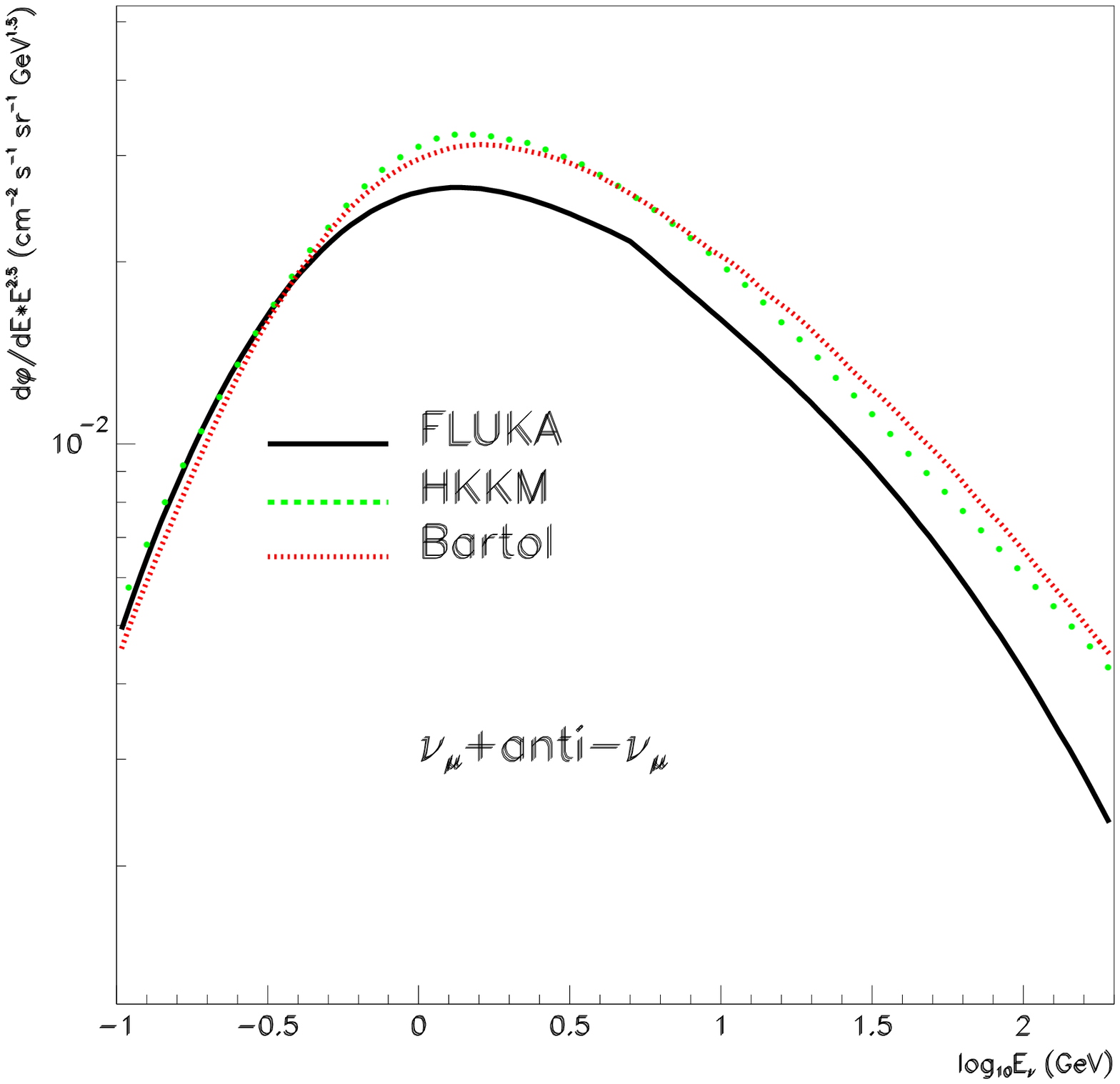,width=75mm}} &
\makebox[75mm]{\epsfig{file=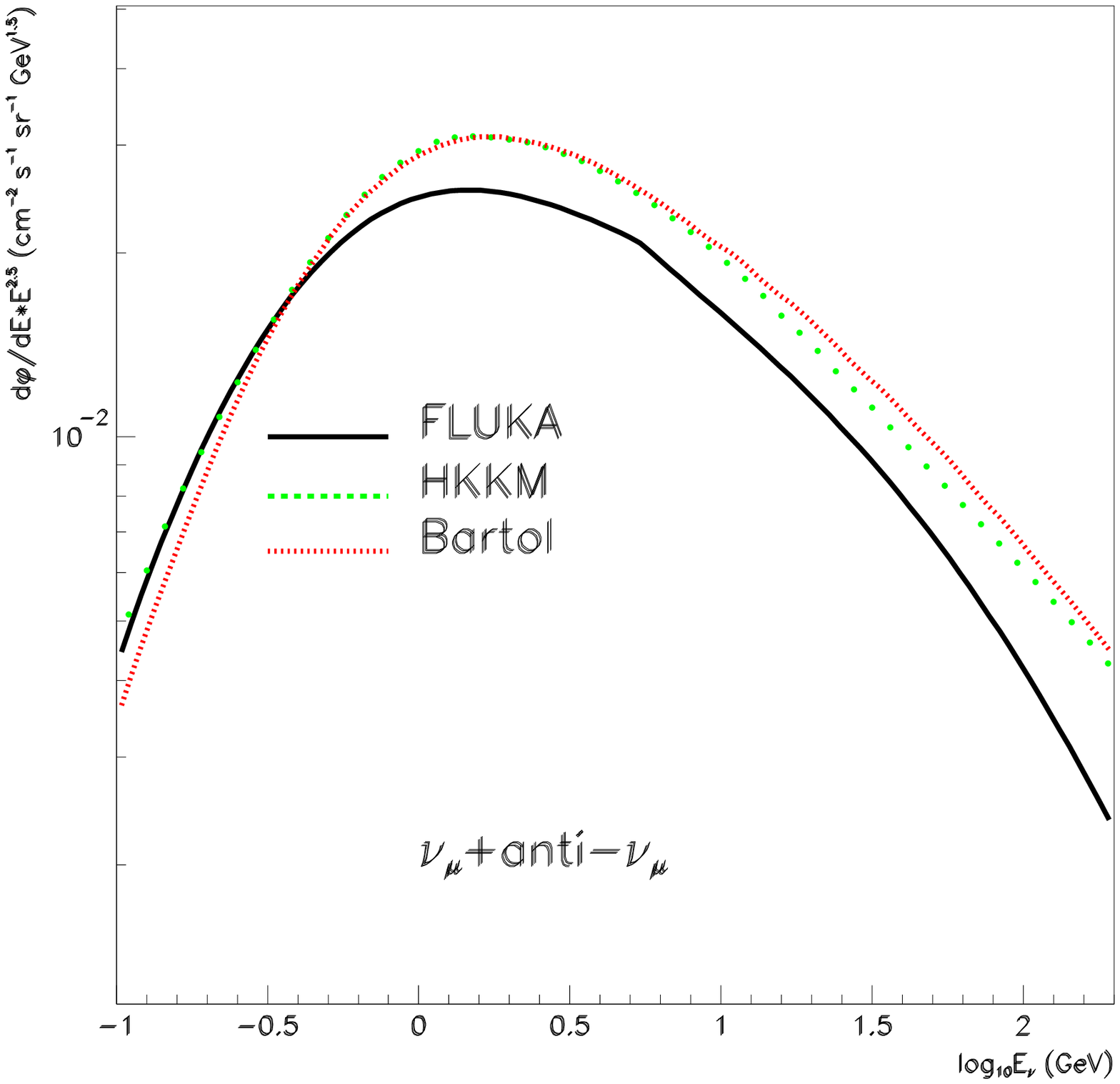,width=75mm}} \\
\end{tabular}
\caption{Muon neutrino and anti-neutrino differential flux times E$^{2.5}$,
 angle averaged over the whole solid angle, 
versus neutrino energy at Super-Kamiokande site for solar minimum (left)
and solar maximum (right)
for the FLUKA calculation (solid line), the Bartol one
(dotted line) \protect\cite{bartol} and for HKKM \protect\cite{hkkm}
 (dashed line).
\label{fluxconfm}
}
\end{center}
\end{figure}

Fig.~\ref{fluxconfm} shows the same comparison but for
muon neutrinos and anti-neutrinos.
It has to be mentioned that both Bartol and the Japanese group
are working to improve their predictions. A part from the
update of the primary flux compilation, Bartol is also improving
their interaction model, as preliminary described in ref.\cite{target},
while Honda and co-workers are now using DPMJET-III\cite{dpmjet3}, and
their new results are reported in \cite{review,hondanew}. There they also
introduce the 3--D geometry, and the geomagnetic field
is included, although in the dipole approximation, also during
shower development. These new results, included the Bartol one,
predict (at least for the Super--Kamiokande site) a lower flux
normalization, now closer to the FLUKA one, with respect
to the previous references of \cite{bartol} and \cite{hkkm}.

\begin{figure}[htbp]
\begin{center}
\makebox[150mm]{\epsfig{file=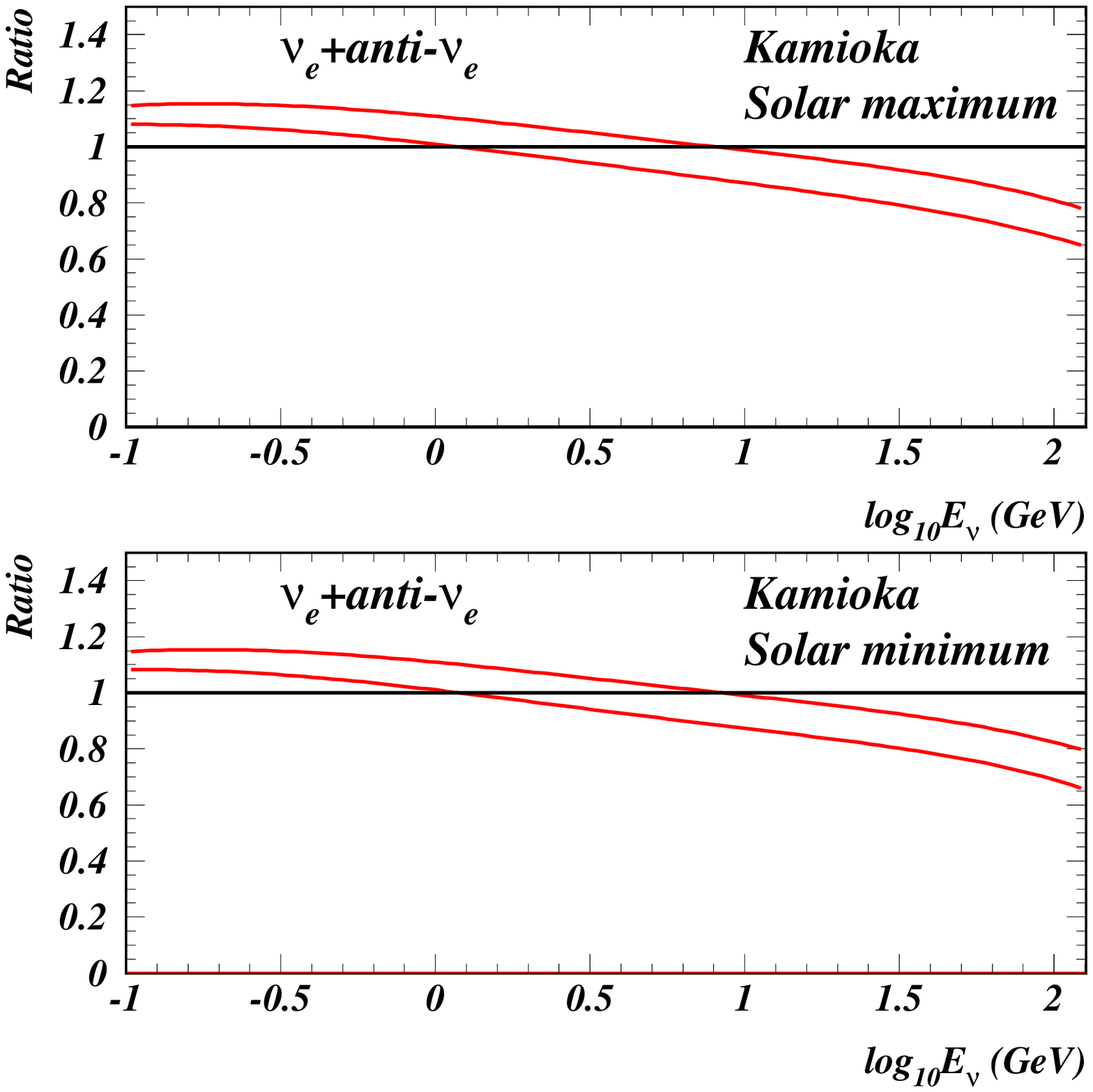,width=150mm%
%,bbllx=60pt,bblly=100pt,bburx=590pt,bbury=740pt}}
,height=15.cm}}
\caption{Muon neutrino and anti-neutrino ratio of fluxes averaged over solid  
angle versus neutrino energy at Super-Kamiokande site and for solar maximum
and solar minimum. The ratio is done between the FLUKA calculation 
using the recent fit to new measurements of primary cosmic rays over
the same calculation using the previous primary
cosmic ray flux used in Ref.\protect\cite{bartol}. The 2 lines forming a 
band represent the uncertainty on the new fit of primaries. 
\label{rappmsk}
}
\end{center}
\end{figure}

\begin{figure}[htbp]
\begin{center}
\makebox[150mm]{\epsfig{file=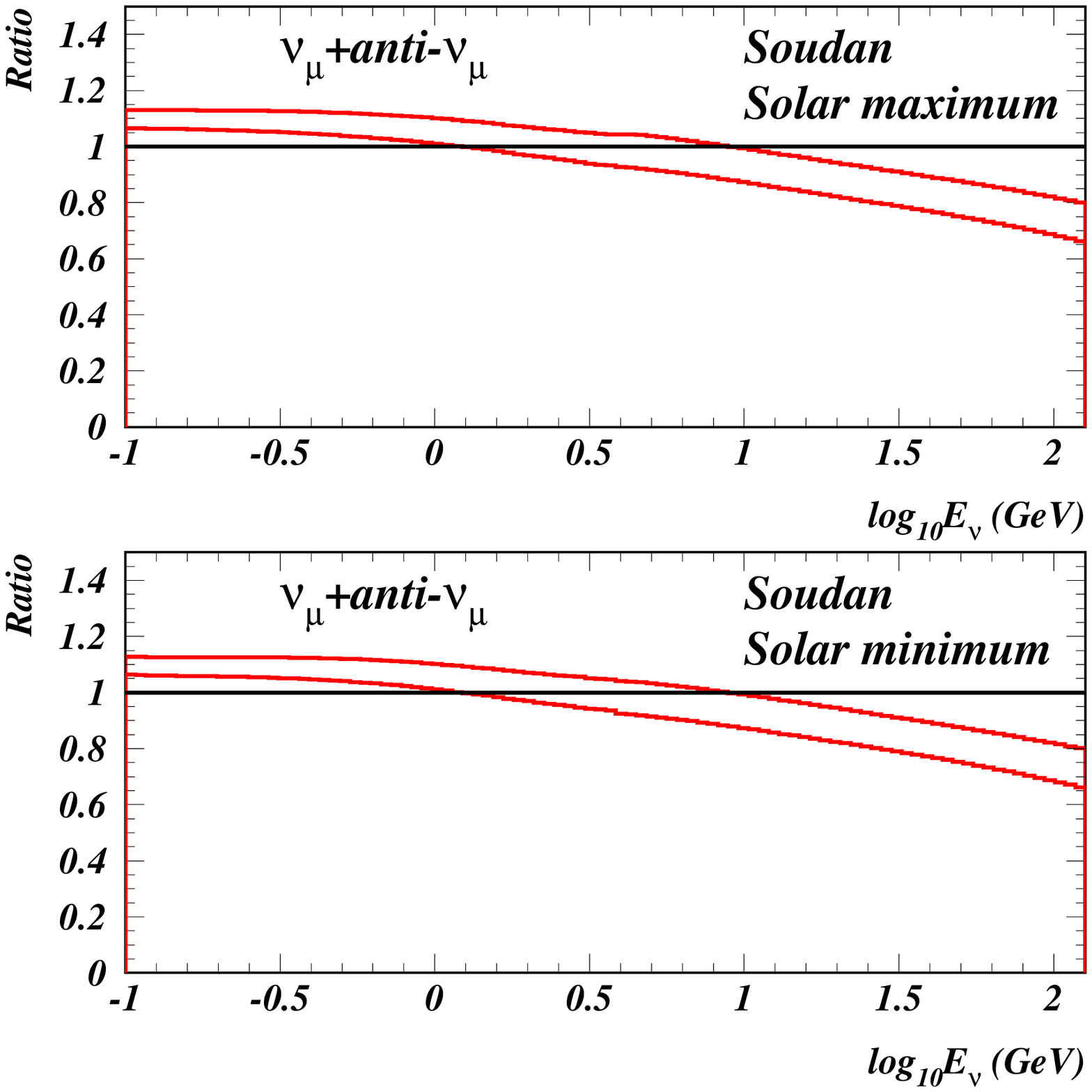,width=150mm%
%,bbllx=60pt,bblly=100pt,bburx=590pt,bbury=740pt}}
,height=15.cm}}
\caption{Muon neutrino and anti-neutrino ratio of fluxes averaged over solid  
angle versus neutrino energy at Soudan site and for solar maximum
and solar minimum. The ratio is done between the FLUKA calculation 
using the recent fit to new measurements of primary cosmic rays over
the same calculation using the previous primary
cosmic ray flux used in Ref.\protect\cite{bartol}. The 2 lines framing a 
band represent the uncertainty on the new fit of primaries. 
\label{rappmsd}
}
\end{center}
\end{figure}

In Fig.~\ref{rappmsk} we show the quantity $R = \Phi(\mbox{new CR
  flux})/\Phi(\mbox{old CR flux})$ calculated 
making the ratio of the solid angle averaged FLUKA muon neutrino plus
antineutrino flux as a function of neutrino energy
for the calculation reported here ({\i.e.} using the
new fit to recent primary cosmic ray measurements) over the flux
as resulting from our previous calculation, where the same primary flux as
  in Ref.~\cite{bartol} was used.
The ratio $R$ is shown at Super-Kamiokande site and for solar maximum and minimum.
The band shows the region between the maximum and minimum fit which can be
obtained from the measurements used.
In  Fig.~\ref{rappmsd} the same ratios are shown for Soudan (low cut-off) site.
The same plots for electron neutrinos are not shown because they are very
similar to those for muon neutrinos.
Notice the ratio represents the effect of changing the primary cosmic ray
spectrum in a neutrino flux calculation, because no other main ingredient
was changed between the newer and older results.
It can be noticed that the improved estimate of the primary cosmic flux 
will increase predictions of the order of 15\%
for the Sub-GeV events, while for Multi-GeV and
through-going muons new predictions should be about 20\% lower in 
normalization.

\begin{center}
\begin{sidewaystable}[p]

\begin{tabular}{|r||c|c|c|c|c|c|c|c|c|c|c|c|}
\hline 
\multicolumn{1}{|c||}{$\phi (\nu_e)$ 
%(cm$^{2}$ s sr GeV)$^{-1}$
}  
& \multicolumn{12}{|c|}{ Cosine of zenith angle} \\
\hline
 E$_\nu$ (GeV)  &
 -0.975 &
 -0.825 &
 -0.625 &
 -0.425 &
 -0.225 &
 -0.025 &
 0.025 &
 0.225 &
 0.425 &
 0.625 &
 0.825 &
 0.975  \\
\hline 
\hline
  0.106&0.25    &0.28    &0.31    &0.30    &0.40    &0.67    &0.67    &0.29    &0.21    &0.18    &0.17    &0.17     \\
  0.168&0.16    &0.18    &0.19    &0.18    &0.22    &0.35    &0.35    &0.17    &0.13    &0.12    &0.11    &0.11     \\
  0.299&0.71E-01&0.77E-01&0.82E-01&0.76E-01&0.86E-01&0.12    &0.12    &0.70E-01&0.60E-01&0.56E-01&0.53E-01&0.51E-01 \\
  0.531&0.24E-01&0.26E-01&0.28E-01&0.26E-01&0.29E-01&0.38E-01&0.38E-01&0.25E-01&0.23E-01&0.21E-01&0.19E-01&0.19E-01 \\
  0.944&0.62E-02&0.67E-02&0.74E-02&0.75E-02&0.83E-02&0.99E-02&0.99E-02&0.79E-02&0.70E-02&0.62E-02&0.56E-02&0.53E-02 \\
  1.679&0.13E-02&0.14E-02&0.16E-02&0.18E-02&0.21E-02&0.24E-02&0.24E-02&0.21E-02&0.18E-02&0.15E-02&0.13E-02&0.12E-02 \\
  2.985&0.21E-03&0.24E-03&0.30E-03&0.36E-03&0.46E-03&0.56E-03&0.56E-03&0.47E-03&0.37E-03&0.29E-03&0.24E-03&0.21E-03 \\
  5.309&0.34E-04&0.40E-04&0.50E-04&0.67E-04&0.95E-04&0.12E-03&0.12E-03&0.95E-04&0.67E-04&0.50E-04&0.39E-04&0.33E-04 \\
  9.441&0.45E-05&0.52E-05&0.70E-05&0.10E-04&0.16E-04&0.24E-04&0.24E-04&0.16E-04&0.10E-04&0.70E-05&0.52E-05&0.45E-05 \\
 16.788&0.57E-06&0.68E-06&0.91E-06&0.14E-05&0.23E-05&0.41E-05&0.41E-05&0.23E-05&0.14E-05&0.91E-06&0.68E-06&0.57E-06 \\
 29.854&0.70E-07&0.85E-07&0.11E-06&0.17E-06&0.32E-06&0.64E-06&0.64E-06&0.32E-06&0.17E-06&0.11E-06&0.85E-07&0.70E-07 \\
 53.088&0.80E-08&0.10E-07&0.14E-07&0.20E-07&0.41E-07&0.90E-07&0.90E-07&0.41E-07&0.20E-07&0.14E-07&0.10E-07&0.80E-08 \\
 94.406&0.85E-09&0.11E-08&0.16E-08&0.23E-08&0.51E-08&0.11E-07&0.11E-07&0.51E-08&0.23E-08&0.16E-08&0.11E-08&0.84E-09 \\
167.880&0.79E-10&0.10E-09&0.19E-09&0.25E-09&0.62E-09&0.12E-08&0.12E-08&0.62E-09&0.25E-09&0.19E-09&0.11E-09&0.81E-10\\
\hline
\end{tabular}
\caption{Electron neutrino flux (cm$^{2}$ s sr GeV)$^{-1}$) as a function of energy for different values
of the cosine of zenith angle as calculated for the Super--Kamiokande site
for maximum solar modulation.
\label{tabe}}
\end{sidewaystable}

\begin{sidewaystable}[p]
\begin{tabular}{|r||c|c|c|c|c|c|c|c|c|c|c|c|}
\hline 
\multicolumn{1}{|c||}{$\phi (\nubar_e)$ 
%(cm$^{2}$ s sr GeV)$^{-1}$
} 
& \multicolumn{12}{|c|}{ Cosine of zenith angle} \\
\hline
 E$_\nu$ (GeV)  &
 -0.975 &
 -0.825 &
 -0.625 &
 -0.425 &
 -0.225 &
 -0.025 &
 0.025 &
 0.225 &
 0.425 &
 0.625 &
 0.825 &
 0.975  \\
\hline 
\hline
  0.106&0.23    &0.25    &0.27    &0.27    &0.36    &0.65    &0.65    &0.28    &0.20    &0.18    &0.17    &0.16     \\
  0.168&0.14    &0.15    &0.16    &0.16    &0.19    &0.32    &0.32    &0.16    &0.12    &0.11    &0.11    &0.10     \\
  0.299&0.62E-01&0.65E-01&0.70E-01&0.65E-01&0.75E-01&0.11    &0.11    &0.64E-01&0.55E-01&0.51E-01&0.49E-01&0.47E-01 \\
  0.531&0.20E-01&0.21E-01&0.23E-01&0.22E-01&0.25E-01&0.33E-01&0.33E-01&0.23E-01&0.20E-01&0.19E-01&0.17E-01&0.16E-01 \\
  0.944&0.51E-02&0.54E-02&0.61E-02&0.62E-02&0.70E-02&0.84E-02&0.84E-02&0.68E-02&0.60E-02&0.54E-02&0.48E-02&0.45E-02 \\
  1.679&0.10E-02&0.11E-02&0.13E-02&0.14E-02&0.17E-02&0.20E-02&0.20E-02&0.17E-02&0.15E-02&0.12E-02&0.10E-02&0.96E-03 \\
  2.985&0.17E-03&0.19E-03&0.23E-03&0.28E-03&0.37E-03&0.45E-03&0.45E-03&0.38E-03&0.29E-03&0.23E-03&0.19E-03&0.17E-03 \\
  5.309&0.27E-04&0.31E-04&0.39E-04&0.50E-04&0.74E-04&0.93E-04&0.93E-04&0.74E-04&0.50E-04&0.38E-04&0.31E-04&0.26E-04 \\
  9.441&0.35E-05&0.42E-05&0.54E-05&0.74E-05&0.13E-04&0.18E-04&0.18E-04&0.13E-04&0.74E-05&0.54E-05&0.42E-05&0.35E-05 \\
 16.788&0.45E-06&0.56E-06&0.71E-06&0.10E-05&0.20E-05&0.34E-05&0.34E-05&0.20E-05&0.10E-05&0.71E-06&0.56E-06&0.45E-06 \\
 29.854&0.54E-07&0.67E-07&0.87E-07&0.13E-06&0.26E-06&0.58E-06&0.58E-06&0.26E-06&0.13E-06&0.87E-07&0.67E-07&0.54E-07 \\
 53.088&0.62E-08&0.69E-08&0.10E-07&0.15E-07&0.32E-07&0.90E-07&0.90E-07&0.32E-07&0.15E-07&0.10E-07&0.69E-08&0.62E-08 \\
 94.406&0.64E-09&0.61E-09&0.12E-08&0.16E-08&0.38E-08&0.12E-07&0.12E-07&0.38E-08&0.16E-08&0.12E-08&0.61E-09&0.64E-09 \\
167.880&0.60E-10&0.46E-10&0.14E-09&0.16E-09&0.44E-09&0.15E-08&0.15E-08&0.44E-09&0.16E-09&0.14E-09&0.46E-10&0.60E-10\\ \hline
\end{tabular}
\caption{Electron anti--neutrino flux (cm$^{2}$ s sr GeV)$^{-1}$) as a function of energy for different values
of the cosine of zenith angle as calculated for the Super--Kamiokande site
for maximum solar modulation.
\label{tabae}}
\end{sidewaystable}

\begin{sidewaystable}[p]
\begin{tabular}{|r||c|c|c|c|c|c|c|c|c|c|c|c|}
\hline 
\multicolumn{1}{|c||}{$\phi( \nu_\mu)$ 
%(cm$^{2}$ s sr GeV)$^{-1}$
} 
& \multicolumn{12}{|c|}{ Cosine of zenith
angle} \\
\hline
 E$_\nu$ (GeV)  &
 -0.975 &
 -0.825 &
 -0.625 &
 -0.425 &
 -0.225 &
 -0.025 &
 0.025 &
 0.225 &
 0.425 &
 0.625 &
 0.825 &
 0.975  \\
\hline 
\hline
  0.106&0.51    &0.56    &0.62    &0.62    &0.81    & 1.5    & 1.5    &0.60    &0.43    &0.37    &0.35    &0.34     \\
  0.168&0.33    &0.35    &0.38    &0.35    &0.43    &0.71    &0.71    &0.34    &0.26    &0.24    &0.23    &0.23     \\
  0.299&0.15    &0.15    &0.16    &0.14    &0.16    &0.24    &0.24    &0.14    &0.12    &0.11    &0.11    &0.11     \\
  0.531&0.51E-01&0.52E-01&0.54E-01&0.49E-01&0.53E-01&0.69E-01&0.69E-01&0.48E-01&0.44E-01&0.42E-01&0.41E-01&0.41E-01 \\
  0.944&0.14E-01&0.14E-01&0.15E-01&0.14E-01&0.15E-01&0.18E-01&0.18E-01&0.15E-01&0.14E-01&0.13E-01&0.13E-01&0.12E-01 \\
  1.679&0.33E-02&0.33E-02&0.35E-02&0.35E-02&0.37E-02&0.43E-02&0.43E-02&0.38E-02&0.36E-02&0.33E-02&0.32E-02&0.31E-02 \\
  2.985&0.68E-03&0.70E-03&0.74E-03&0.78E-03&0.86E-03&0.97E-03&0.97E-03&0.89E-03&0.80E-03&0.74E-03&0.70E-03&0.68E-03 \\
  5.309&0.14E-03&0.15E-03&0.16E-03&0.17E-03&0.20E-03&0.21E-03&0.21E-03&0.19E-03&0.16E-03&0.15E-03&0.14E-03&0.14E-03 \\
  9.441&0.26E-04&0.27E-04&0.28E-04&0.32E-04&0.37E-04&0.41E-04&0.41E-04&0.37E-04&0.32E-04&0.28E-04&0.27E-04&0.26E-04 \\
 16.788&0.48E-05&0.51E-05&0.51E-05&0.60E-05&0.71E-05&0.76E-05&0.76E-05&0.71E-05&0.60E-05&0.52E-05&0.51E-05&0.48E-05 \\
 29.854&0.85E-06&0.89E-06&0.93E-06&0.11E-05&0.13E-05&0.14E-05&0.14E-05&0.13E-05&0.11E-05&0.94E-06&0.89E-06&0.85E-06 \\
 53.088&0.14E-06&0.14E-06&0.16E-06&0.18E-06&0.22E-06&0.24E-06&0.24E-06&0.22E-06&0.18E-06&0.16E-06&0.14E-06&0.14E-06 \\
 94.406&0.20E-07&0.21E-07&0.24E-07&0.28E-07&0.34E-07&0.39E-07&0.39E-07&0.34E-07&0.28E-07&0.24E-07&0.21E-07&0.20E-07 \\
167.880&0.27E-08&0.27E-08&0.33E-08&0.41E-08&0.47E-08&0.60E-08&0.60E-08&0.46E-08&0.42E-08&0.33E-08&0.27E-08&0.27E-08\\ \hline
\end{tabular}
\caption{Muon neutrino flux (cm$^{2}$ s sr GeV)$^{-1}$) as a function of energy for different values
of the cosine of zenith angle as calculated for the Super--Kamiokande site
for maximum solar modulation.
\label{tabm}}
\end{sidewaystable}

\begin{sidewaystable}[p]
\begin{tabular}{|r||c|c|c|c|c|c|c|c|c|c|c|c|}
\hline 
\multicolumn{1}{|c||}{$\phi_{\nubar_\mu}$ 
%(cm$^{2}$ s sr GeV)$^{-1}$
} 
& \multicolumn{12}{|c|}{ Cosine of zenith
angle} \\
\hline
 E$_\nu$ (GeV)  &
 -0.975 &
 -0.825 &
 -0.625 &
 -0.425 &
 -0.225 &
 -0.025 &
 0.025 &
 0.225 &
 0.425 &
 0.625 &
 0.825 &
 0.975  \\
\hline 
\hline
  0.106&0.51    &0.56    &0.62    &0.60    &0.81    & 1.4    & 1.4    &0.60    &0.43    &0.38    &0.35    &0.35     \\
  0.168&0.33    &0.35    &0.38    &0.35    &0.43    &0.69    &0.69    &0.34    &0.26    &0.24    &0.23    &0.23     \\
  0.299&0.15    &0.15    &0.16    &0.14    &0.16    &0.24    &0.24    &0.14    &0.12    &0.11    &0.11    &0.11     \\
  0.531&0.50E-01&0.51E-01&0.53E-01&0.49E-01&0.52E-01&0.68E-01&0.68E-01&0.47E-01&0.43E-01&0.42E-01&0.41E-01&0.40E-01 \\
  0.944&0.13E-01&0.14E-01&0.14E-01&0.14E-01&0.15E-01&0.18E-01&0.18E-01&0.14E-01&0.13E-01&0.13E-01&0.12E-01&0.12E-01 \\
  1.679&0.30E-02&0.31E-02&0.33E-02&0.33E-02&0.36E-02&0.42E-02&0.42E-02&0.37E-02&0.34E-02&0.31E-02&0.29E-02&0.28E-02 \\
  2.985&0.59E-03&0.62E-03&0.67E-03&0.72E-03&0.84E-03&0.97E-03&0.97E-03&0.86E-03&0.74E-03&0.67E-03&0.62E-03&0.59E-03 \\
  5.309&0.12E-03&0.12E-03&0.13E-03&0.15E-03&0.18E-03&0.21E-03&0.21E-03&0.18E-03&0.14E-03&0.13E-03&0.12E-03&0.12E-03 \\
  9.441&0.20E-04&0.22E-04&0.24E-04&0.27E-04&0.33E-04&0.45E-04&0.45E-04&0.33E-04&0.27E-04&0.24E-04&0.22E-04&0.20E-04 \\
 16.788&0.35E-05&0.38E-05&0.42E-05&0.48E-05&0.58E-05&0.86E-05&0.86E-05&0.58E-05&0.48E-05&0.43E-05&0.38E-05&0.35E-05 \\
 29.854&0.60E-06&0.65E-06&0.72E-06&0.82E-06&0.10E-05&0.15E-05&0.15E-05&0.10E-05&0.81E-06&0.71E-06&0.65E-06&0.60E-06 \\
 53.088&0.97E-07&0.11E-06&0.11E-06&0.13E-06&0.17E-06&0.23E-06&0.23E-06&0.17E-06&0.13E-06&0.11E-06&0.11E-06&0.97E-07 \\
 94.406&0.15E-07&0.16E-07&0.15E-07&0.20E-07&0.28E-07&0.31E-07&0.31E-07&0.29E-07&0.20E-07&0.15E-07&0.17E-07&0.15E-07 \\
167.880&0.21E-08 &0.23E-08&0.19E-08&0.29E-08&0.45E-08&0.38E-08&0.38E-08&0.47E-08&0.30E-08&0.19E-08&0.24E-08&0.21E-08\\ \hline
\end{tabular}
\caption{Muon anti--neutrino flux (cm$^{2}$ s sr GeV)$^{-1}$) 
as a function of energy for different values
of the cosine of zenith angle as calculated for the Super--Kamiokande site
for maximum solar modulation.
\label{tabam}}
\end{sidewaystable}
\end{center}

\section{Estimate of Systematic Errors}
\label{sec:syserr}
Uncertainties in the flux of
atmospheric neutrinos are presently attributed 
mostly to the primary spectrum
and to the treatment of hadronic interactions. 
In addition, there are minor uncertainties
related to the details of shower 
calculation (atmosphere, technical algorithms, etc.).

As far as the primary spectrum is concerned, trusting the
uncertainty on the fit of \cite{bartol_new}, we derive from
Fig.~\ref{rappmsk} and \ref{rappmsd} that this translates
into an uncertainty of about $\pm$7\%, up to 100 GeV.

The attribution of a systematic error to the hadronic
interaction model is difficult, since there is not
the possibility of comparing predictions to experimental
data in the whole phase space of interest.
The existing data point out that, on average, 
the agreement of \FLUKA{} predictions of data is at the level
of about 10\%. In some limited region of phase space
the level of agreement might be worse, although not above 20\%. 
In the recent years, a lot of work has been done instead to compare
different models to establish a sort of relative uncertainty.
If we consider in the comparison the first models adopted by
Bartol and Honda et al., differences of the order of $\pm$20 to 25\%
have been found, at least for the range of energy
considered in this paper. In our opinion this is not to
be considered as a correct estimate of the overall theoretical uncertainty,
but instead a measurement of the insufficient accuracy of the
first models. Indeed, the recent developments
and upgrades are converging towards a narrower uncertainty
interval, which we estimate not larger than $\pm$10 to 15\%.
The Sub-GeV region of atmospheric neutrinos seems the one
where the theoretical uncertainty appears to be dominant, and 
this reflects the poor knowledge of particle productions in
nucleon--Nucleus interaction at few GeV's.
Recently, we have also made preliminary investigations of
the possible error introduced by assuming the validity
of the superposition model\cite{taup2001}, by testing a preliminary interface to
DPMJET-II.5\cite{dpmjet25} inside FLUKA to perform the
explicit nucleus-nucleus Nucleus-Air interaction. The results
do not show any significant deviation from those obtained
with the superposition model, in contrast to what instead
is claimed in ref.\cite{Fior}.

In any case, the predictions on flavor ratio, which is an
essential element for the analysis of neutrino data in terms
of new physics (oscillations), have always been much less
dependent on the hadronic model: $\pm$2 to 5\%.
As a first approximation, the main source of uncertainty in the prediction
of the standard-model $\mu/e$  ratio comes from
the difference between $\pi^+$ and $\pi^-$.
 The decay of $\pi^+$ eventually results in $\nu_e$ while the decay of $\pi^-$
generates $\nubar_e$; 
in both cases a pair ($\nu_\mu$,$\nubar_\mu$) is also produced.

About other sources of uncertainties, the most relevant
seem to be the choice of an average atmosphere and the common practice
of neglecting mountain shapes for the prediction in specific sites.
As far as the atmosphere is concerned, we agree that
pressure and density variations may have a noticeable impact
on muon fluxes, and mostly at ground level. For neutrinos
these turn out to be much less important, since the effect
is averaged over many directions, and we estimated
a contribution not larger that 1\% (up to 100 GeV of neutrino
energy).

In our calculation we have not yet introduced the geomagnetic
field during shower development in atmosphere. From a set of small scale tests, we
expect that this might have brought to an overestimate of the
absolute flux by few percent. Furthermore, we also know that
this choice would produce a non correct east--west asymmetry,
as compared with data.

The previous considerations can be used to provide an overall
estimate of the theoretical uncertainty of the present calculation.
Taking 7\% for the primary spectrum, 15\% for the interaction model,
1\% for the atmosphere and 2\% for the geomagnetic field and
combining them in quadrature, a total of 17\% is obtained.

As a last remark, we want to to state that, as far as the
comparison with experimental data on atmospheric neutrinos are concerned,
another important source of uncertainty must be included, that is
the one associated to neutrino--Nucleus cross section.

\section{Conclusions}
\label{sec:concl}
The first phase of atmospheric neutrino flux calculation
using the \FLUKA{} code is here concluded.
We think that these results, apart from the question of
the 3-D geometry, assessed already in \cite{flukanu},
represent the first systematic attempt to explore the
impact of a refined hadronic interaction model, which is capable
at the same time to reproduce a wide range of accelerator and
cosmic ray data.
This work has stimulated a serious debate inside the scientific
community about the validity of the previous ``traditional'' calculations,
and we consider the recent convergence towards a lower normalization
of neutrino fluxes (for the same or similar primary spectrum) as 
an important achievement and recognition of the importance of
using accurate interaction models for cosmic ray calculations.
Along this line other attempts have been followed, like the 
works of ref.\cite{Tserk,Fior,Wentz,Liu,Plyaskin}.
Some of these attempts are, in our opinion, biased again by 
the choice of interaction models which are not precise.
In particular we refer to those which are based upon the use of the old 
GHEISHA model\cite{gheisha},
a parametrized code which fails in giving proof
of reliability in reproducing particle production properties,
as recently shown for instance in the framework of
ALICE experiment at LHC\cite{alice}.  
The calculation of
ref.\cite{Tserk}, which is based on GEANT-FLUKA, cannot be compared
to ours, since the FLUKA package contained in GEANT-3 is 
an old and incomplete version of the present FLUKA. In particular
it does not contain the fundamental PEANUT section and the
high energy part (above 5 GeV) is now considered obsolete.
The work of ref.\cite{Plyaskin} is instead originally biased
by a technical error in the normalization, as recently
communicated by the author\cite{plya2}.

As mentioned above, these results cannot yet be considered
as completely final, since we are aware that a fully certified
calculation must include the geomagnetic field also during shower
development. However this has to be done using the most accurate
description of this field, avoiding approximations, and this will
be the object of our next development.
Furthermore, the calculation of neutrino fluxes is also
going to be extended at higher energies, as soon as it will be
released a next extension of the
\FLUKA{} model in order to deal with nucleus--nucleus interactions up
to extreme high energies\cite{flukaV}. 

The importance of reducing as much as possible the theoretical
uncertainties in the calculation of these fluxes may have
limited impact in the analysis of present experimental results
concerning neutrino oscillations where normalization is left as a free
parameter so that results are not dependent on it.
The matter can be different in the framework of a 3--flavor scenario,
where sub-GeV electron neutrinos acquire some weight: there in fact, 
one expects to see the effects of interference terms 
involving $\theta_{12}$, if the LMA solution for solar neutrino
turns out to be the right one, as confirmed by the recent SNO results\cite{sno}.
From the experimental point of view, the ICARUS experiment could be
the one who can investigate with low or negligible systematic error
the sector of low energy electron neutrinos in the atmospheric flux.

\section*{Acknowledgments}
We wish to thank Prof. C. Rubbia, the Icarus collaboration
and the MACRO collaboration for the strong support
to this effort. We acknowledge the fruitful discussions 
with R. Engel, T.K. Gaisser, M. Honda, 
T. Kajita, P. Lipari, V. Naumov, T. Stanev and F. Villante.
The Bartol group provided us the primary spectrum fit.
The Super--Kamiokande collaboration, through Dr. Choji Saji, 
provided us with the HKKM tables of angle integrated fluxes.

\clearpage

\clearpage

\begin{thebibliography}{9}
\bibitem{hkkm2} The HKKM fluxes tabulated for different modulations
have been provided by the Super--Kamiokande group (Dr. Choji Saji).
\bibitem{flukaV} This extension is under way as a 
part of the program of the current development of \FLUKA{}  inside INFN.

\newpage
\end{thebibliography}
\end{document}